\documentstyle[12pt,aaspp4,epsf]{article}

\begin{document}

\def\Ibardd{\skew6 \ddot{I\mkern-6.8mu\raise0.3ex\hbox{-}}}
\def\Ibar{I\mkern-6.8mu\raise0.3ex\hbox{-}}

\title{Newtonian Hydrodynamics of the Coalescence of Black Holes with
Neutron Stars I: Tidally locked binaries with a stiff equation of
state.}

\author{William H. Lee\altaffilmark{1} \& W\l odzimierz Klu\'{z}niak} 

\affil{University of Wisconsin--Madison, Physics Department, \\
1150 University Ave., Madison, WI, 53706 \\
Copernicus Astronomical Center, ul. Bartyczka 18, 00-716 Warszawa, Poland} 

\altaffiltext{1}{Present address: \\ 
Instituto de Astronom\'{\i}a--UNAM \\
Apartado Postal 70--264 \\
Cd. Universitaria \\
M\'{e}xico DF 04510 \\
MEXICO \\
wlee@astroscu.unam.mx
}

\lefthead{Lee \& Klu\'{z}niak}
\righthead{Black Hole--Neutron Star coalescence}

\begin{abstract}

We present a detailed study of the hydrodynamical interactions in a
Newtonian black hole--neutron star binary during the last stages of
inspiral. We consider close binaries which are tidally locked, use a
stiff equation of state (with an adiabatic index $\Gamma=3$)
throughout, and explore the effect of different initial mass ratios on
the evolution of the system. We calculate the gravitational radiation
signal in the quadrupole approximation. Our calculations are carried
out using a Smooth Particle Hydrodynamics~(SPH) code. 

\end{abstract}

Subject headings: binaries: close --- gamma rays: bursts ---
hydrodynamics --- stars: neutron

\section{Introduction}

It is our objective in the present work to investigate the outcome of
the coalescence of a black hole with a neutron star, to find out to
what extent the neutron star is tidally disrupted, and in particular
to determine if an accretion structure does form around the black hole
as a result of the encounter. We also wish to explore the role of the
mass ratio and the stiffness of the equation of state in the evolution
of the system and in the emission of the gravitational radiation. For
simplicity and to perform accurate comparisons with the work presented
by~\cite{RS94}, (hereafter RS) for double neutron star binaries, we
have chosen to model the neutron star as a polytrope and explore
different mass--radius relationships by varying the stiffness through
the adiabatic exponent $\Gamma$. In this paper we restrict ourselves
to a stiff equation of state with $\Gamma$=3.

We restrict our work to binaries which are tidally locked. The
construction of a completely self--consistent initial condition is
then entirely straightforward. This assumption represents a major
simplification in our calculations, and can be seen as an extreme case
of angular momentum distribution in the system. The opposite extreme
would be a case where the binary components exhibit no rotation at all
as viewed from an external, inertial frame of reference, and it poses
a much greater problem regarding the construction of self--consistent
initial conditions in equilibrium. Since complete tidal locking is not
expected~(\cite{bild}), the full range of initial configurations
deserves to be explored.

The motivation for our work is presented in
section~\ref{motivation}. The numerical method we have used to carry
out our simulations is described in section~\ref{method}. Calibration
to previous work and our modeling of the black hole--neutron star
system is presented in section~\ref{calibration}. In
section~\ref{results} we present our numerical results, and a
discussion of these results follows in section~\ref{discussion}.

\section{Motivation \label{motivation}}

The orbital evolution of short--period binary systems containing
compact objects (black holes or neutron stars) will inevitably lead to
coalescence due to orbital angular momentum loss to gravitational
radiation. Such binary systems have been found (PSR
1913+16---\cite{hulse}, PSR 1534+12---\cite{wol}) and the rate of
inspiral matches the prediction of general relativity to high
accuracy~(\cite{taylor,taylorb}). The time to final merging in these
binaries is less than the Hubble time and so they will eventually
coalesce. In the final stages of such a coalescence, a powerful burst
of gravitational waves is expected, and as such these systems are
primary candidate sources for detection by instruments such as LIGO
and VIRGO, expected to begin operation within a few
years~(\cite{snowmass}). Estimates for the event rate can be inferred
from the statistics of the known Hulse--Taylor type binaries, and a
rate of $10^{-6}$ to $10^{-5}$ per galaxy per year can be
expected~(\cite{latt,nara}), which would imply several coalescences
per year could be observed out to a distance of 1~Gpc. These
gravitational waves will certainly carry a copious amount of
information about the emitting source. Among other things, they are
expected to exhibit strong departure from point--mass behavior due to
hydrodynamical effects and the finite size of the stars, as well as a
dependence on the details of the equation of state for matter at
nuclear densities.

A fully three--dimensional hydrodynamical study is therefore essential
to understand the gravitational radiation signal, and this information
can be used to place constraints on the equation of state. For the
typical masses and binary separations involved in very close binary
systems (a few solar masses and several tens of kilometers
respectively), one should ideally solve the problem using
general--relativistic formalism and include radiation processes. This
has not yet been realized, but some qualitative results can hopefully
be obtained from a Newtonian treatment.

It has been shown~(\cite{LRSa}) that at least in the Newtonian case,
hydrodynamic effects play an important role in the orbital evolution
of the system. Essentially, tidal interactions can make a close binary
dynamically unstable for small enough separations---on the order of a
few stellar radii---when the spin angular momentum becomes comparable
to the orbital angular momentum. This effect alone can produce orbital
decay on a time scale comparable to that of angular momentum loss to
gravitational waves.

The final stages of inspiral, and the coalescence of double neutron
star binaries has been the object of numerous previous studies in a
Newtonian framework. The gravitational radiation signal was studied by
Nakamura \& Oohara~(\cite{nakamuraI}) and Oohara \&
Nakamura~(\cite{nakamuraII},~\cite{nakamuraIII}) while Davies {\em et
al.}~(\cite{Davies}) considered the effect of different initial spin
configurations on the outcome of the merger. More recent simulations
have calculated the neutrino emission as well as the gravitational
waves from the coalescence (Ruffert {\it et
al.}~\cite{ruffI,ruffII,ruff}). The de--stabilization effect of tidal
forces in a Newtonian framework has been investigated in great
detail~(\cite{LRSb}, RS) using an analytical approach based on an
energy variational method to examine equilibrium configurations and a
three--dimensional numerical treatment to study the hydrodynamical
aspects of the coalescence, using throughout a polytropic equation of
state.

There is yet another reason why these systems are the focus of intense
study, since they have been suggested~(\cite{eich}) as a possible
source for the production of gamma--ray bursts (GRBs). It is well
established that the GRB distribution on the sky is isotropic and it
is generally believed that their sources lie at cosmological
distances~(\cite{fish}), with redshifts on the order of
unity~(\cite{metzger}). The observed fluence of said bursts,
$10^{-7.5}$ to $10^{-3}$~erg~cm$^{-2}$, and the rise times in their
light curves imply that an energy release of at least $10^{51}$~erg
takes place in a few milliseconds, in a region at most a few hundred
kilometers across.

To produce a GRB, this amount of energy, after being released in a
primary event, must be transformed into $\gamma$--rays. The
relativistic blast wave model~(\cite{mesz,rees}) requires a relatively
baryon--free line of sight to the observer along which matter can be
accelerated to speeds close to the speed of light. The interaction of
this outflow with the interstellar medium would then presumably lead
to shock acceleration of electrons and emission of $\gamma$--rays
through synchrotron radiation.

In the double neutron star merger scenario~(\cite{bpap,eich}) the
merger initially produces a burst of neutrinos and
anti--neutrinos. However, Newtonian calculations~(\cite{janka}) have
shown that the energy release into neutrinos is insufficient to power
the observed GRBs. General relativistic calculations carried out in a
conformally flat metric~(\cite{wilson}) have shown that in some cases
each neutron star may collapse into a black hole many orbits prior to
merging so that no blast wave will occur. Nevertheless, this could
still produce a GRB with a smooth time profile~(\cite{wilson97}). In
the black hole--neutron star merger scenario~(\cite{bp}), it was
expected that the neutron star will be disrupted by the black hole and
a thick accretion torus will be formed. As in the case with two
neutron stars, this could lead to the formation of a blast wave which
would power the GRB.

\section{Numerical Method \label{method}}

For the calculations presented in this paper, we have used the
numerical technique known as Smooth Particle Hydrodynamics~(SPH). This
is essentially a Lagrangian method, where forces are evaluated by
interpolation over a grid of points co--moving with the fluid, which
can be considered as particles. This method was originally developed
by~\cite{Lucy} and \cite{GM} as an alternative to Eulerian
hydrodynamics (computations on a fixed grid). The principal advantages
of SPH are that no assumptions need to be made {\it a priori} about
the nature of the flow that will be studied and that no computational
effort is wasted in modeling regions where matter is not present. This
is particularly advantageous when studying complicated
three--dimensional astrophysical flows. An excellent review of the
method has been given by~\cite{Monaghan92}.

SPH has been tested successfully and applied to a variety of problems,
such as the shock wave in a tube problem~(\cite{MG}), redistribution
of angular momentum in a thick accretion torus~(\cite{ZB}), static
stellar structure~(\cite{GM}), astrophysical jets~(\cite{CB}), and
hydrodynamics of close binary systems~(\cite{Benz}, \cite{RS92},
\cite{Davies}). We have developed our own SPH code~(\cite{phd}) and
successfully tested it in one, two and three dimensions with several
of these problems. Agreement with previous results has been excellent
in every case. A description of our code is given in
Appendix~\ref{code}.

\section{Calibration and initial conditions \label{calibration}}

\subsection{Simulation of a double neutron star binary \label{nsns}}

Given that the problem of merging black hole--neutron star binaries is
closely related to the study of coalescing double neutron star or
white dwarf binaries, we have performed a rigourous calibration of our
code to the results presented by RS for two polytropes with stiff
equations of state ($\Gamma$=3) and an initial mass ratio of unity in
a tidally locked binary. While our simulations had a lower resolution
($\sim$2000 particles per star vs. $40000$ for RS), qualitative and
quantitative agreement was excellent.

For the following, measure distance and mass in units of the radius
and mass of the unperturbed (spherical) neutron star (13.4~km and
1.4$M_{\odot}$ respectively), except where noted, so that the units of
time, density and velocity are:
\begin{eqnarray}
\tilde{t}= 1.146 \times 10^{-4}~{\rm s} \times
\left(\frac{R}{13.4~\mbox{km}}\right)^{3/2}
\left(\frac{M}{1.4M_{\odot}}\right)^{-1/2}
\label {eq:deftunit}
\end{eqnarray}
\begin{eqnarray}
\tilde{\rho}= 1.14 \times 10^{18}~{\rm kg~m}^{-3} \times
\left(\frac{R}{13.4~\mbox{km}}\right)^{-3}
\left(\frac{M}{1.4M_{\odot}}\right)
\label{eq:defrhounit}
\end{eqnarray}
\begin{eqnarray}
\tilde{v}=0.39c \times \left(\frac{R}{13.4~\mbox{km}}\right)^{-1/2}
\left(\frac{M}{1.4M_{\odot}}\right)^{1/2}
\label{eq:defvunit}
\end{eqnarray}
where $R$ and $M$ are the radius and mass of the unperturbed
(spherical) neutron star.

The coalescence resulted in a massive central core containing about
82\% of the total mass, surrounded by a massive halo and extended
spiral arms. At the end of the calculation, the central object is not
azimuthally symmetric and thus continues to emit gravitational waves,
of amplitude $h_{final}$. The amplitudes and frequencies of the
gravitational radiation waveforms were found to be in excellent
agreement with RS (see~\cite{LK},~\cite{phd}). In Table~\ref{RSLK} we
present a comparison of some of the more important parameters in the
case of the coalescence of two neutron stars. The first, second and
third columns show the maximum and final amplitude in the
gravitational radiation waveforms and the maximum gravitational wave
luminosity respectively. These values are given in geometrized units,
where $G$=$c$=1, and the peak luminosity $L_{max}$ is normalized to
$L_{0}$=$c^{5}/G$=$3.59\times 10^{59}$~erg~s$^{-1}$. The fourth and
fifth columns display the central density of the resulting core and
the mass contained in the halo surrounding it, as a fraction of the
total mass in the system.

For the simulations of a black hole--neutron star binary presented
here, we have chosen to keep a stiff equation of state ($\Gamma$=3) to
compare our results to the case of the coalescence of two neutron
stars. In future work, we will explore different values of the
adiabatic index $\Gamma$ to model different mass--radius
relationships. This is particularly important since future
measurements of the gravitational radiation emitted during these
events could serve to constrain the equation of state for dense
matter.

\subsection{Modeling of the black hole \label{black}}

In our simulations, the black hole is modeled by a point mass, and
produces a Newtonian potential:
\begin{eqnarray*}
\Phi_{BH}(r)=-\frac{GM_{BH}}{r}. \label{eq:bhphi}
\end{eqnarray*}
The contribution from the black hole, to the force on particle {\em i}
in the star is then simply given by:
\begin{eqnarray*}
\mbox{\boldmath $F^{BH}_{i}$} =-\frac{GM_{BH}m_{i}}{ \mid
\mbox{\boldmath$ r_{i}-r_{BH}$} \mid ^{3} } ( \mbox{\boldmath
$r_{i}-r_{BH}$} ) , \label{eq:Fibh}
\end{eqnarray*}
and symmetrically,
\begin{eqnarray*}
\mbox{\boldmath $F^{i}_{BH}$} =-\frac{Gm_{i}M_{BH}}{ \mid
\mbox{\boldmath$ r_{i}-r_{BH}$} \mid ^{3} } ( \mbox{\boldmath
$r_{BH}-r_{i}$} ) , \label{eq:Fbhi}
\end{eqnarray*}
is the contribution from particle {\em i} to the force on the black hole. 

The horizon of the black hole is modeled by placing an absorbing
boundary at a distance $r_{s}$=$2GM_{BH}/c^{2}$ from the point
mass. At every time step during the dynamical simulations, any
particle that crosses this boundary is absorbed by the black hole and
removed from the simulation. The mass, position and velocity of the
black hole are adjusted so that total mass and linear momentum are
conserved. We thus disregard any spin angular momentum that might be
gained by the black hole during the process. This does not present any
problems, since our calculation is Newtonian throughout, and we make
no attempt to model frame--dragging.

\subsection{Construction of initial conditions \label{init}}

Our simulations begin with the construction of a spherical,
unperturbed polytrope, which for simplicity we will refer to as the
neutron star. To do this, we proceed essentially as Rasio \& Shapiro
(1992). A total of $N$ particles are placed on a cubic lattice with
masses $m_{i}$=$\rho_{LE}/n$, where $\rho_{LE}$ is the density
calculated from the Lane--Emden solution to the equation of
hydrostatic equilibrium with an equation of state $P$=$K\rho^{\Gamma}$
and {\em n} is the number density of particles. The smoothing length
is assigned to each particle so that the number of overlapping
neighbors per particle is $\nu$=64. This amounts to setting the
smoothing length $h_{i}\sim l$, where {\em l} is the lattice
spacing. There is considerable advantage in having variable particle
masses, as this increases the spatial and density resolution near the
edge of the star, where the density gradient is largest. The radius of
the unperturbed polytrope is 13.4~km, and its mass is
$1.4M_{\odot}$. For the simulations presented in this paper, we have
used $N \sim 8000$ for every case except two, where $N \sim 17000$.

To obtain tidally locked equilibrium configurations for the binary
system, the unperturbed polytrope and the black hole are placed in the
co--rotating Keplerian frame of initial angular velocity $\Omega$=
$\sqrt{G(M+M_{BH})/r^{3}}$, where {\em r} is the binary
separation defined as the distance between the black hole and the
center of mass of the neutron star. A damping term linear in the
velocity is introduced into the equations of motion for the SPH
particles to allow the star to respond to the presence of the tidal
gravitational field. We now have:
\begin{eqnarray*}
m_{i} \dot{\mbox{\boldmath$v$}}_{i}=\mbox{\boldmath $F_{iG}+F_{iH}
-v_{i}$}/t_{damp} + m_{i}\Omega^{2} \mbox{\boldmath $r_{i}$} ,
\label{eq:damping}
\end{eqnarray*}
\begin{eqnarray*}
M_{BH} \dot{\mbox{\boldmath$v$}}_{BH}=\mbox{\boldmath $F_{BH}+$}
M_{BH}\Omega^{2} \mbox{\boldmath $r_{BH}$} .  \label{eq:dampingbh}
\end{eqnarray*}
where \mbox{\boldmath $F_{iG}$} and \mbox{\boldmath $F_{iH}$} are the
gravitational and hydrodynamical forces on particle {\em i} (of mass
$m_{i}$) respectively. The value of $t_{damp}$ is chosen so that
oscillations are critically damped, reaching an equilibrium as quickly
as possible; \(t_{damp}=(GM/R^{3})^{-\frac{1}{2}}\). We neglect
Coriolis forces since we are interested in equilibrium configurations
with no bulk motion in the co--rotating frame. The position of the
black hole and the center of mass of the star are adjusted at every
time step so that the binary separation remains at a desired value. In
the same manner, $\Omega$ is adjusted so that the total (gravitational
plus centrifugal) force on the black hole in the co--rotating frame,
\mbox{\boldmath $F_{BH}$}$+M_{BH}\Omega^{2}$\mbox{\boldmath $r_{BH}$}
is zero, i.e.:
\begin{eqnarray*}
\Omega = \sqrt{\frac{F_{BH}}{M_{BH}r_{BH}}}, \label{eq:defomega}
\end{eqnarray*}
where \mbox{\boldmath$r_{BH}$} is the position vector of the black
hole. This procedure ensures that the configuration reaches
equilibrium in a state of synchronization.

For a given value of the binary separation, we allow the system to
relax for a period of twenty time units, keeping the specific
entropies of all particles constant, i.e. $K$=constant in
$P$=$K\rho^{\Gamma}$. In all cases, our initial conditions satisfy the
virial ratio to better than three parts in $10^{3}$. We have varied
the mass ratio in the binary, $q$=$M/M_{BH}$, by changing the mass of
the black hole only.

\section{Results \label{results}}

\subsection{Introduction}

In this section we present the results of our simulations for
different values of the mass ratio $q$ in the binary, beginning with
high values of $q$ and proceeding in descending order. The units used
are as defined in equations~(\ref{eq:deftunit}), (\ref{eq:defrhounit})
and (\ref{eq:defvunit}) except where noted. The highest value of the
mass ratio is unity. Although the production of a binary system with
such a low--mass black hole is rather unlikely, we nevertheless wish
to carry out a comparison with the case for two neutron stars, and use
this as a starting point in our investigation. For each mass ratio we
present the initial configurations that were constructed for tidally
locked binaries, followed by a description of the dynamical runs.

To investigate the dynamical evolution of the system for a given
initial separation, we remove the damping term from the equations of
motion and give every SPH particle and the black hole the azimuthal
velocity corresponding to the equilibrium value of $\Omega$ in an
inertial frame, with the origin at the center of mass of the
system. Each SPH particle is assigned a specific thermal energy
$u_{i}$=$K\rho^{(\Gamma-1)}/(\Gamma-1)$, and the equation of state is
changed to that of an ideal gas where $P$=$(\Gamma-1)\rho u$. The
specific thermal energy of each SPH particle is then evolved
individually throughout the simulation according to the first law of
thermodynamics, taking into account the contribution from the viscous
terms (see Appendix~\ref{code}).

\subsection{Mass ratio $q$=1}

\subsubsection{Equilibrium configurations \label{initq1}}

\begin{figure}
\plotone{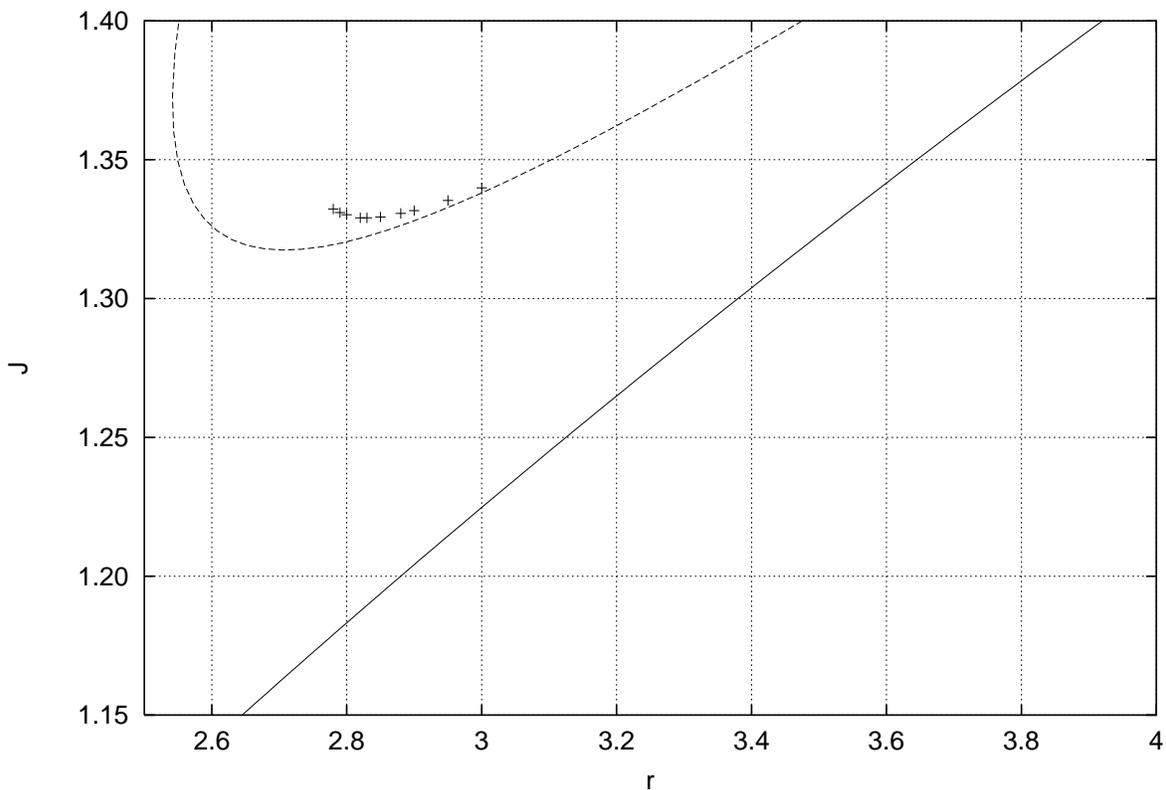}
\caption{Total angular momentum vs. binary separation for a black
hole--neutron star binary with mass ratio $q$=1. The solid line is the
result for two point masses, the dotted line is computed using the
analytical approach of~\cite{LRSb} treating the neutron star as a
compressible tri--axial ellipsoid, and the crosses are the result of
SPH relaxation calculations. The Roche limit is at $r$=2.78.
\label{jvsrq1}}
\end{figure}

\begin{figure}
\plotone{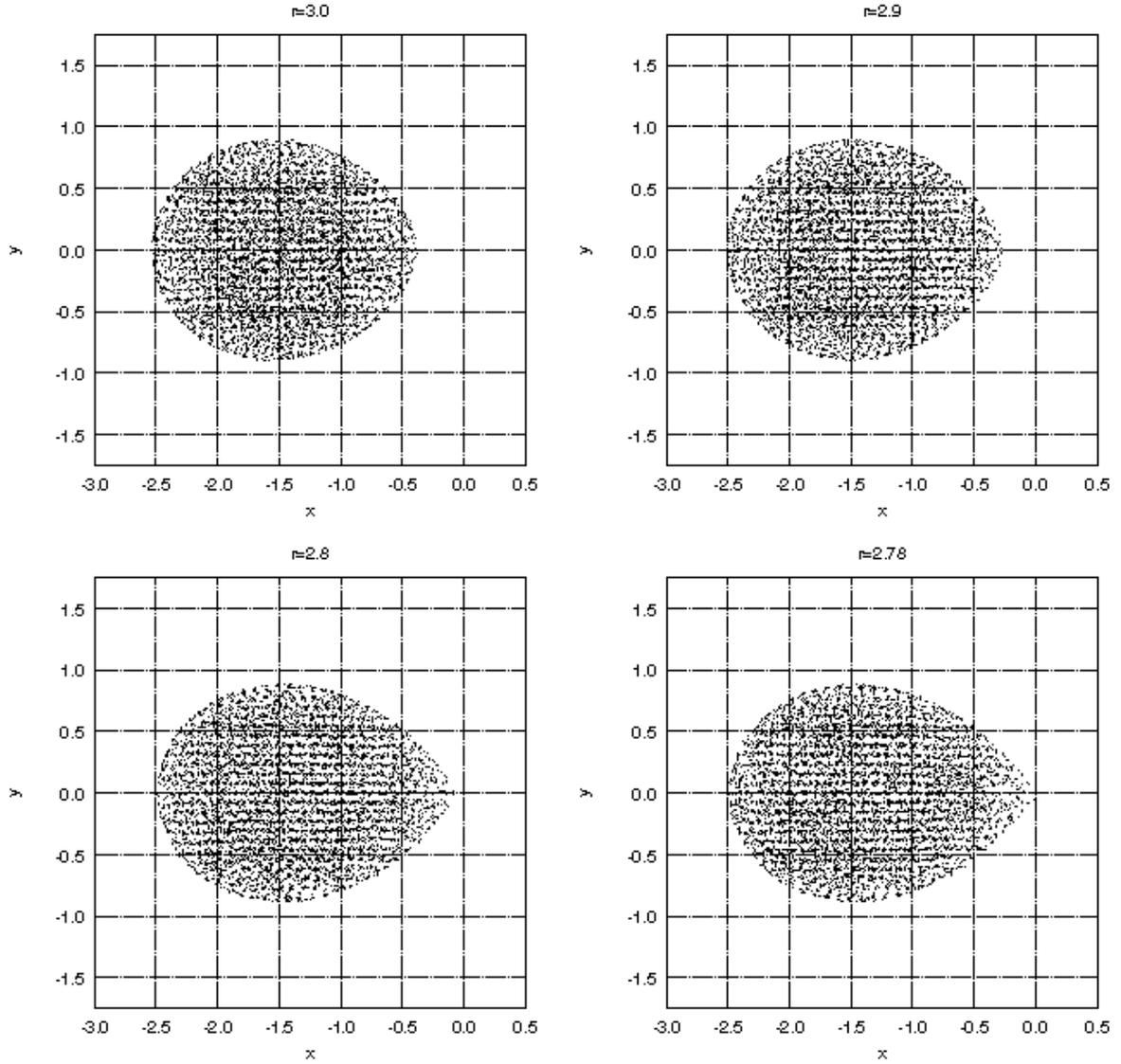}
\caption{Particle positions projected onto the equatorial plane for
initial configurations at various binary separations. One dot is
plotted for every SPH particle.
\label{initposq1}}
\end{figure}

As described above (section~\ref{init}), equilibrium configurations
were constructed for different values of the binary separation {\em
r}. We show in Figure~\ref{jvsrq1} 
a plot of the total angular
momentum of the system, {\em J}, versus binary separation {\em r}. The
solid line is the result for two point masses in Keplerian orbit. The
dotted line is the result of approximating the neutron star as a
compressible tri--axial ellipsoid~(\cite{LRSb}), and one cross is
plotted for each SPH calculation at a fixed separation (with $N$=8121
SPH particles for the neutron star). The presence of a minimum in {\em
J} is a direct result of the assumption of tidal locking of the
extended star. Essentially, as the separation is decreased, the spin
component of angular momentum must grow (just as the orbital component
is decreasing), and eventually become comparable to the orbital
component. The result is that the total angular momentum {\em J}
reaches a minimum and as the separation decreases further, it will
increase. Such a turning point in the curve of total angular momentum
as a function of separation marks the onset of an instability, which
will lead to orbital decay~(\cite{LRSa}). Note the strong departure
from point--mass behavior, and the excellent agreement of the full SPH
calculations with the compressible tri--axial ellipsoid treatment
before the minimum in {\em J} is achieved. Close to the minimum, the
ellipsoidal approximation breaks down and a full numerical treatment
is necessary. The cross with the smallest value of $r$ corresponds to
Roche--Lobe overflow ($r$=$r_{RL}$=2.78).  Our initial determination
of the stability limit~(\cite{LK}) was higher than this value because
our initial conditions did not correspond to proper tidal locking. We
show in Figure~\ref{initposq1} 
plots of particle positions projected
onto the equatorial plane for different separations. As the separation
is decreased, the effect of the tidal potential becomes more apparent
and the neutron star is distorted. The configuration at $r$=3.0 can be
approximated by a tri--axial ellipsoid, while for those at $r$=2.8 and
$r$=2.78 this is increasingly no longer the case.

\subsubsection{Dynamical runs \label{dynq1}}

\begin{figure}
\plotone{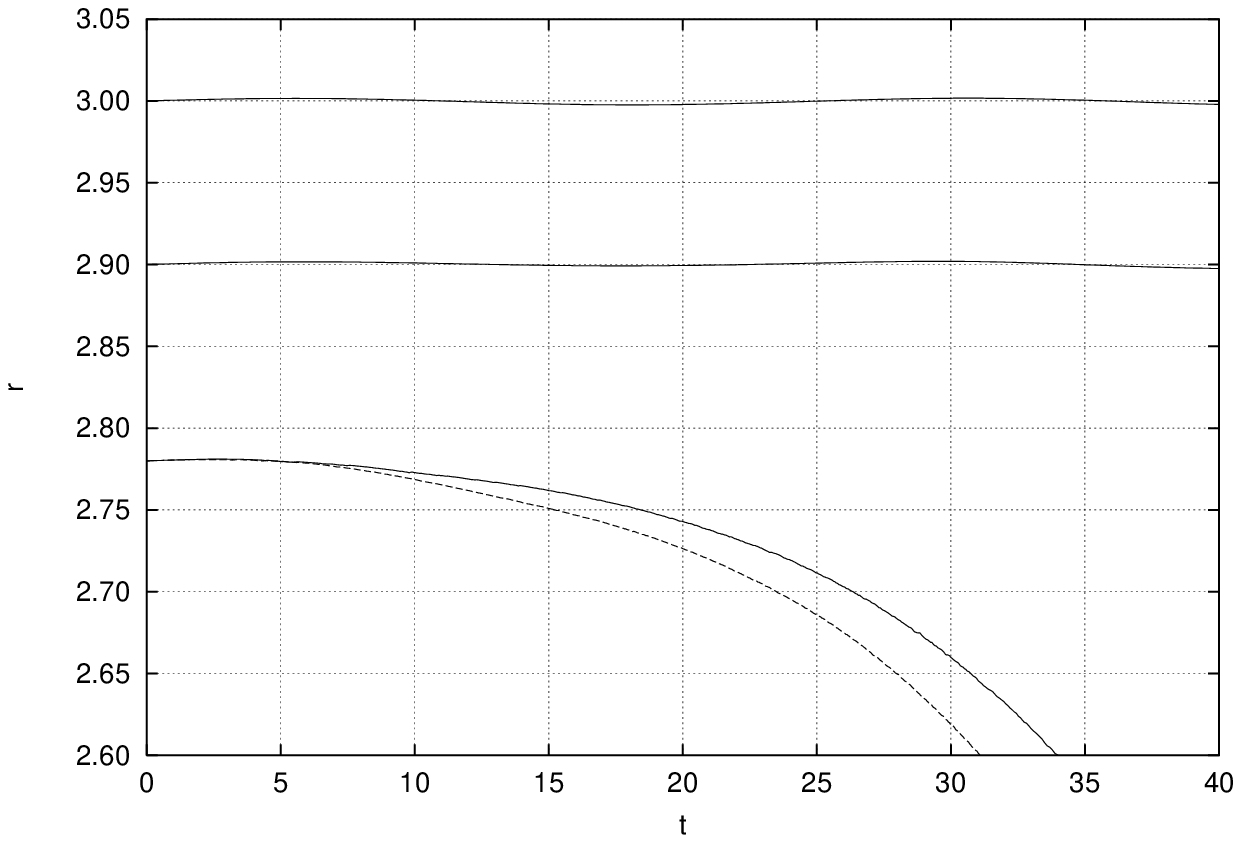}
\caption{Binary separation as a function of time for dynamical
calculations starting at three different separations ($r$=3.0, $r$=2.9
and $r$=2.78) for the black hole--neutron star binary with mass ratio
$q$=1. For $r$=2.78, the solid line used $N$=8121 particles and the
dotted line used $N$=16944 particles. \label{rvstq1}}
\end{figure}

We have used several of the configurations thus constructed to perform
dynamical runs. In Figure~\ref{rvstq1} 
we show the binary separation
as a function of time during several of these calculations.  It is
clear that the configurations with initial separations $r$=3.0 and
$r$=2.9 are stable. The separation exhibits oscillations of numerical
origin on a period close to the orbital period (respectively $P$=22.81
and $P$=21.6 in our units, see eq.~[\ref{eq:deftunit}]). However, the
configuration with $r$=2.78 and initial period $P$=20.09
(corresponding to 37~km and 2.39~ms for $R$=13.4~km and
$M$=$1.4M_{\odot}$) is unstable, and the separation decays on an
orbital timescale. We find the dynamical stability limit to be
$r_{dyn}$=2.78=$r_{RL}$. The two different lines with initial
separation $r$=2.78 correspond to two different resolutions (the solid
line used $N$=8121 particles and the dotted line used $N$=16944
particles). In the following, we will describe the results for our
higher resolution run, with $N$=16944.  In Figure~\ref{rhobhq1} 
\begin{figure}
\plotone{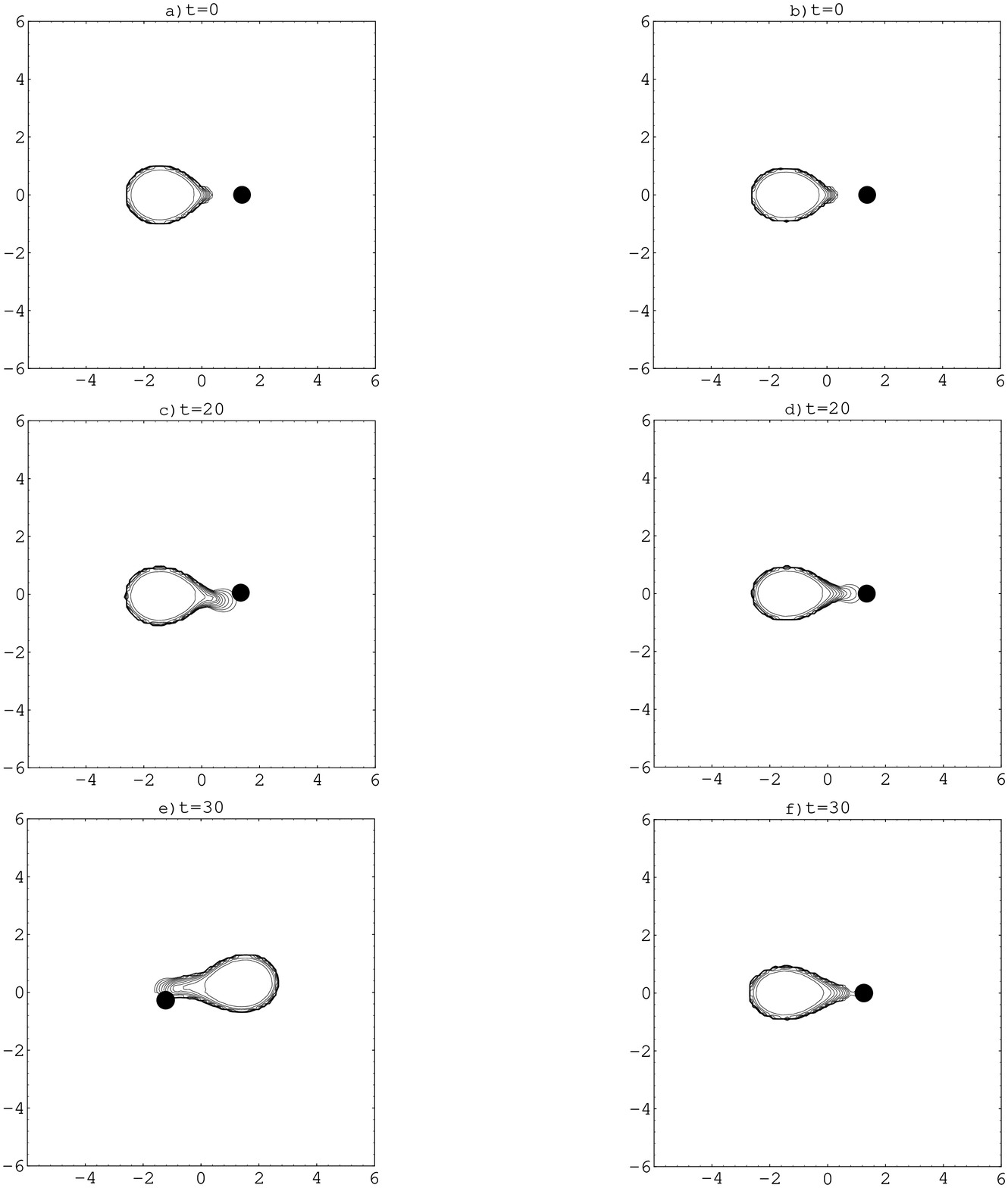}
\caption{Density contours at various times during
the dynamical coalescence of the black hole--neutron star binary with
mass ratio $q$=1 and initial separation $r$=2.78. The logarithmic
contours are evenly spaced every 0.25 decades, with the lowest one at
$\log{\rho}=-3.25$. The rotation is counterclockwise and the initial
orbital period is $P$=20.09. The left column shows density contours in
the orbital plane, the right column shows contours in the meridional
plane containing the binary axis. The black disk represents the black
hole. \label{rhobhq1}}
\end{figure}
\setcounter{figure}{3}
\begin{figure}
\plotone{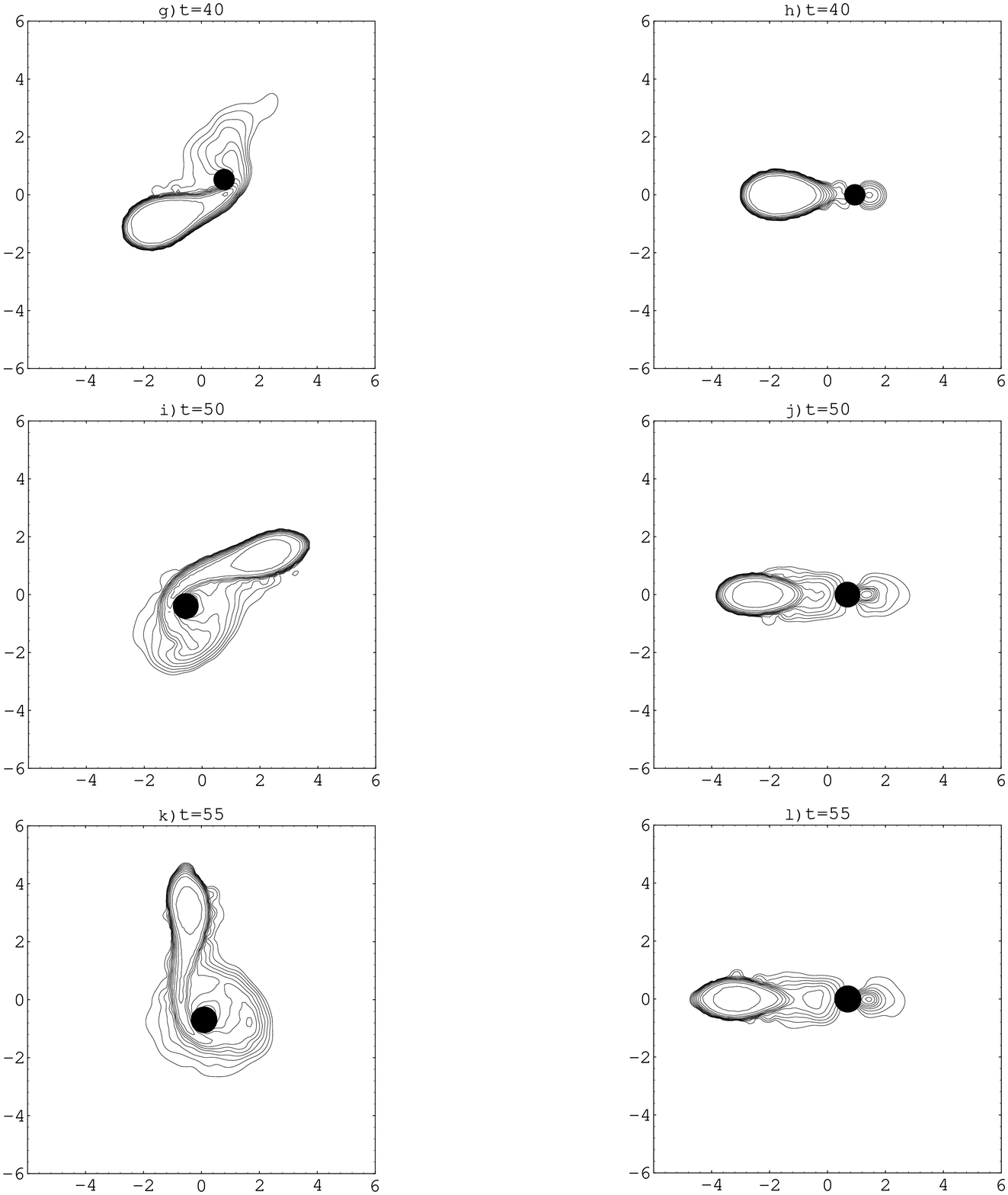}
\caption{continued.}
\end{figure}
\setcounter{figure}{3}
\begin{figure}
\plotone{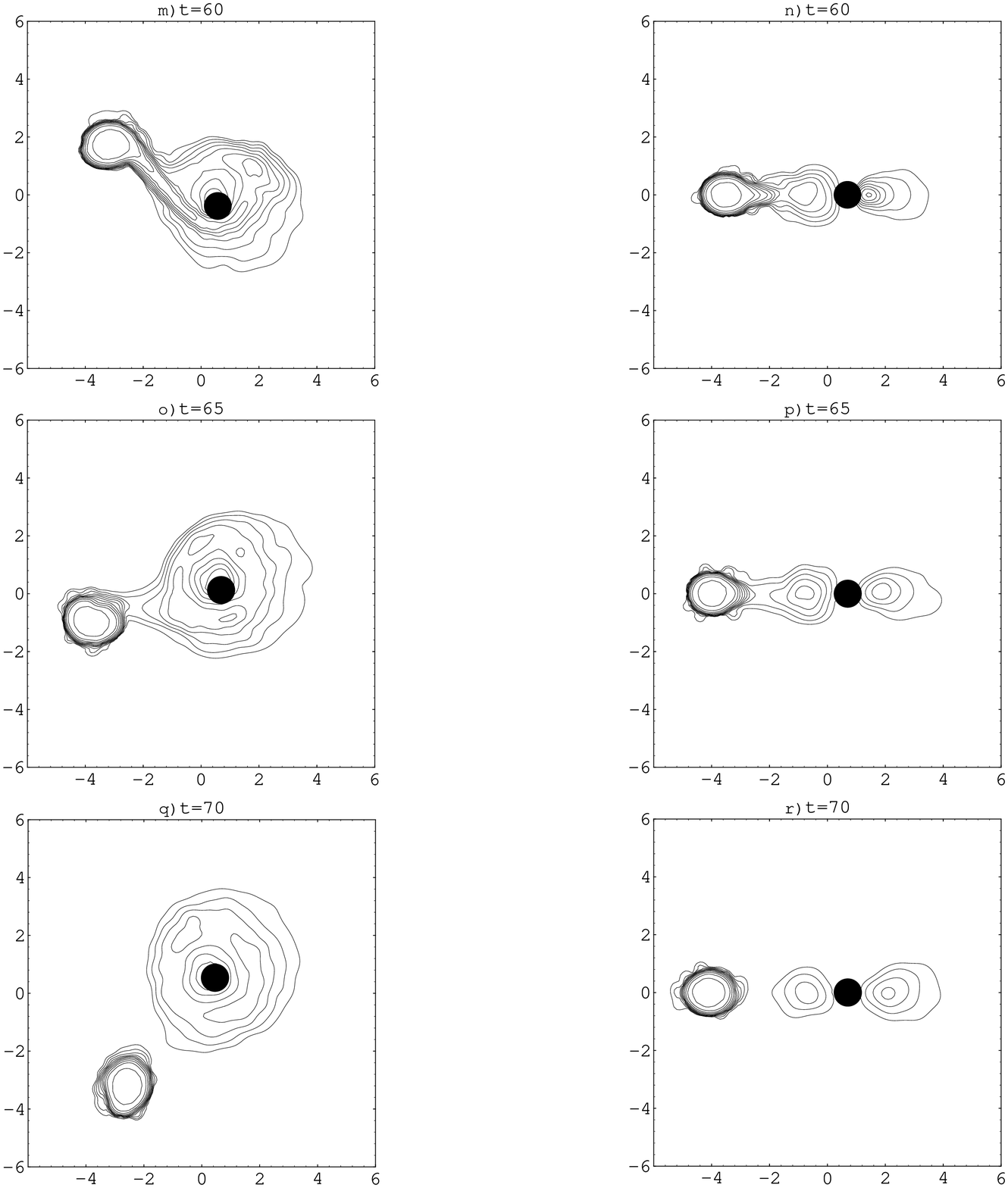}
\caption{continued.}
\end{figure}
\setcounter{figure}{3}
\begin{figure}
\plotone{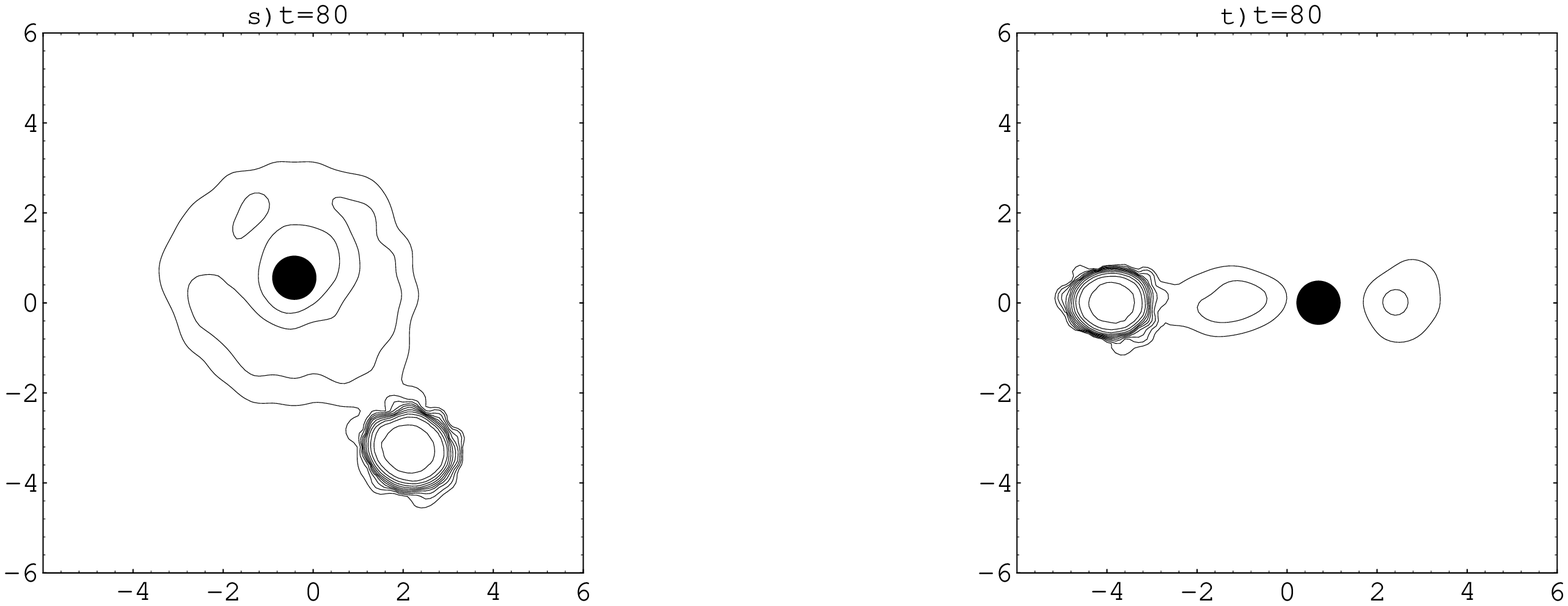}
\caption{continued.}
\end{figure}
we show density contours at various times during the simulation, which
was run from $t$=0 to $t$=100. The left column shows contours in the
orbital plane, and the right column shows contours in the meridional
plane containing the black hole and the center of mass of the SPH
particles. The black disk with radius $r_{s}$ represents the black
hole. Mass transfer from the neutron star to the black hole begins
almost immediately (within one orbit) through the inner Lagrangian
point. The accretion stream gradually becomes thicker and begins to
wrap around the black hole. By $t$=50 the star has become considerably
elongated and a toroidal structure has begun to form around the black
hole. The accretion stream then essentially breaks, mass transfer
stops and a stellar core remains in orbit around the black hole
($t$=60 through $t$=70). The result of this coalescence then appears
to be a a stable binary with a greatly altered mass ratio and
separation ($q_{final}$=0.19 and $r$$\sim$4.5, corresponding to
60~km), consisting of a remnant core with mass
$M_{core}$=0.307~(corresponding to $0.43M_{\odot}$) in orbit around a
black hole with $M_{BH}$=1.607~(or $2.25M_{\odot}$).  The mass
transfer event is very brief, as can be seen in Figure~\ref{mbhq1},
where
\begin{figure}
\plotone{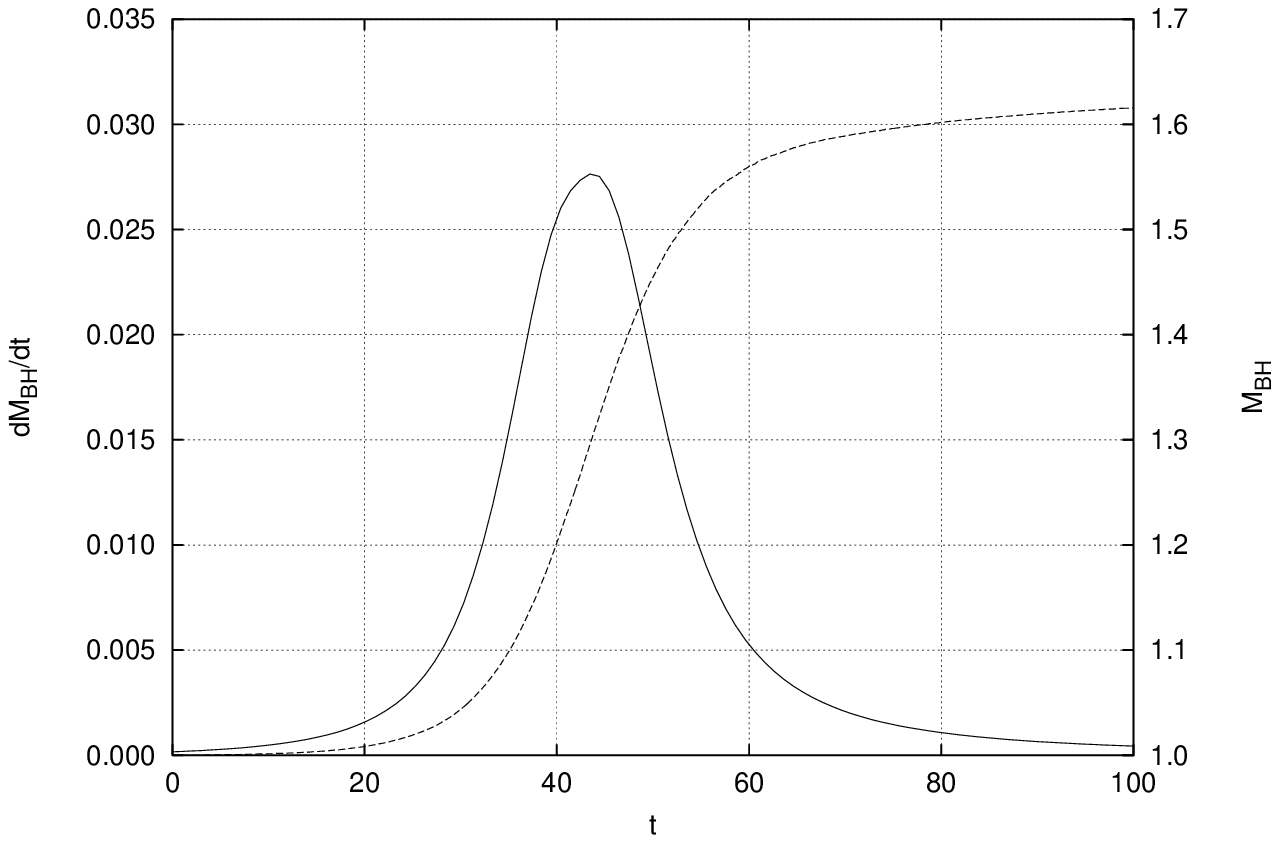}
\caption{Black hole mass (dotted line) and mass accretion rate onto
the black hole (solid line) as a function of time for the black
hole--neutron star binary with mass ratio $q$=1 and initial separation
$r$=2.78. \label{mbhq1}}
\end{figure}
we plot the mass of the black hole and the mass accretion rate
onto the black hole as a function of time. The peak accretion rate is
$dM_{BH}/dt$=0.0275 (corresponding to $0.3M_{\odot}/$ms). The black
hole is still accreting matter towards the end of our simulation, but
the decreased resolution (recall that accretion entails a loss of
particles in the simulation) does not allow us to determine the
distribution of matter accurately beyond $t$=100 ($\sim$11~ms). What
can be determined is that the toroidal structure that forms does {\em
not} extend to form a halo that engulfs the black hole completely. The
region directly above the black hole remains devoid of matter down to
the limit of our resolution of $7.1\times 10^{-5}$~(or
$10^{-4}M_{\odot}$) within $\sim 5^{\rm o}$. The accretion structure
around the black hole was not previously observed~(\cite{LK}) because
of the low resolution ($N$=2176 particles) of our initial runs.

\begin{figure}
\plotone{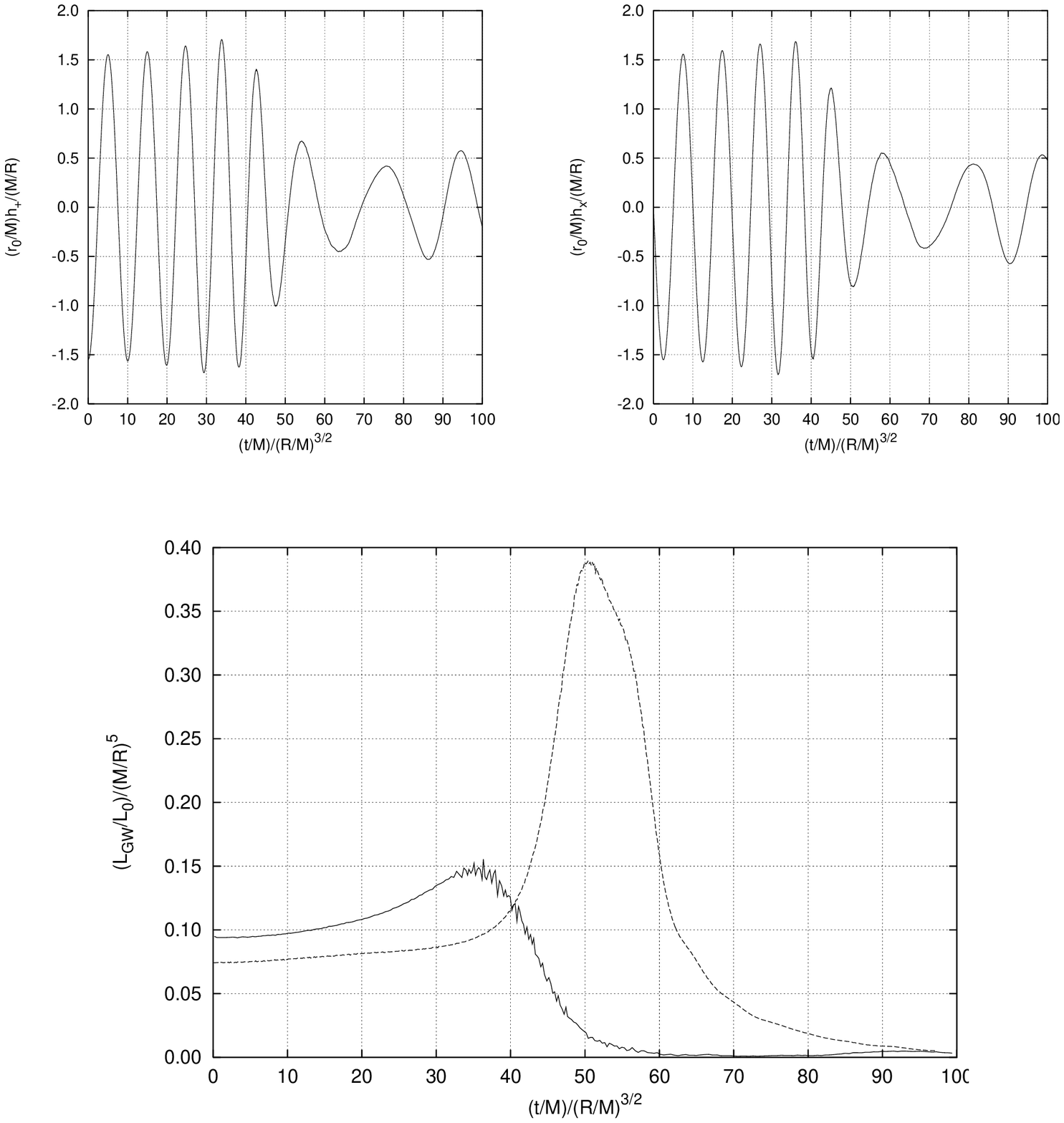}
\caption{Top: Gravitational radiation waveforms for the coalescence of
a black hole--neutron star binary with mass ratio $q$=1. Two
polarizations ($h_{+}$, left panel; $h_{\times}$, right panel) in
geometrized units (such that $G$=$c$=1) are shown. Bottom:
Gravitational radiation luminosity in geometrized units ($G$=$c$=1)
for the coalescence of a black hole--neutron star binary (solid line)
and a double neutron star binary (dotted line) with mass ratio
$q$=1. The constant
$L_{0}$=$c^{5}/G$=$3.6\times10^{59}$~erg~s$^{-1}$.\label{bhnsq1grav}}
\end{figure}

The nature of the encounter is reflected in the gravitational
radiation waveforms, presented in Figure~\ref{bhnsq1grav}~(top).
The amplitude of the emitted waves initially rises from
$(r_{0}R/M^{2})h$=1.55 to $(r_{0}R/M^{2})h_{max}$=1.75, reflecting the
decrease in separation. The subsequent drop in amplitude and decrease
in frequency are a result of the short episode of mass transfer, and
the final waveforms reflect the fact that the binary has survived the
encounter (the new binary orbital period is $P\approx 39$ in our
units, or 4.5~ms). Thus the presence of a remnant core in orbit around
the black hole would be immediately apparent from an observation of
the waveforms emitted during such a coalescence.  We also show the
gravitational wave luminosity in Figure~\ref{bhnsq1grav}~(bottom). For
comparison purposes, the luminosity for the coalescence of two
identical neutron stars is plotted as well, from the results of
section~\ref{nsns}. The maximum luminosity and total energy radiated
in gravitational waves are $(R/M)^{5}L_{max}^{NS-NS}$=0.39 (or
$1.17\times 10^{55}$~erg~s$^{-1}$), $(R^{7/2}/M^{9/2})\Delta
E_{GW}^{NS-NS} \approx 10$ (or $3.4\times 10^{52}$~erg) and
$(R/M)^{5}L_{max}^{BH-NS}$=0.15 (or $4.5\times 10^{54}$~erg~s$^{-1}$),
$(R^{7/2}/M^{9/2})\Delta E_{GW}^{BH-NS} \approx 6$ (or $2\times
10^{52}$~erg) for the double neutron star and black hole--neutron star
case respectively. Although in both cases there is still a clear rise
and decay in the luminosity curve, the peak is much broader ($\Delta
t^{BH-NS}\sim 25$) than for two neutron stars ($\Delta t^{NS-NS}\sim
15 $). The slower rise in luminosity and longer timescale of the peak
in the black hole--neutron star case are a direct consequence of
having only one star being deformed and disrupted through tidal
interactions.

\begin{figure}
\plotone{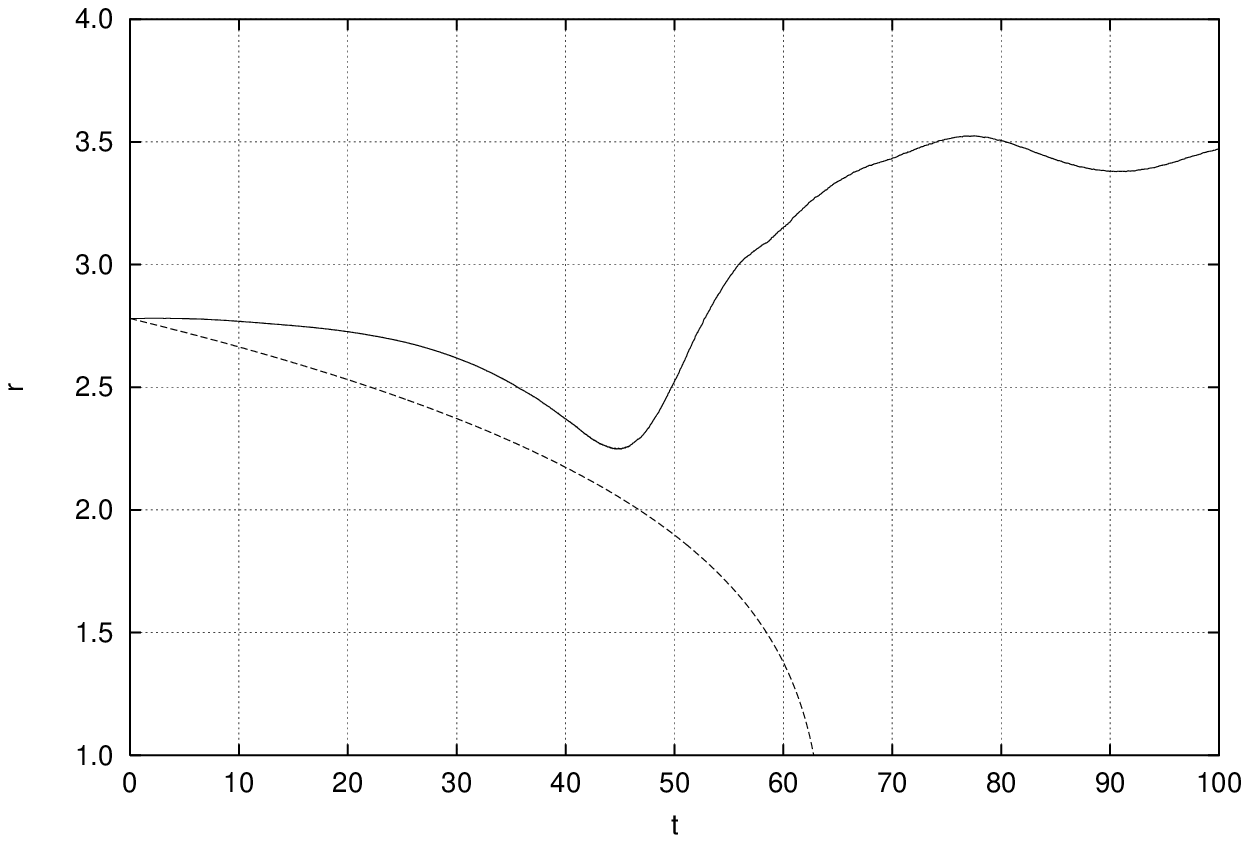}
\caption{Binary separation for the coalescence of a black
hole--neutron star binary with mass ratio $q$=1 (solid line) and for a
point--mass binary with the same mass ratio and initial separation
decaying via the emission of gravitational waves in the quadrupole
approximation (dotted line). \label{ptdecayq1}}
\end{figure}

Finally, we show in Figure~\ref{ptdecayq1} 
the binary separation as a
function of time for the simulation we have just presented, together
with a plot of the separation for a point--mass binary with the same
mass ratio and initial separation decaying via the emission of
gravitational waves in the quadrupole approximation. This last curve
is given by (e.g.~\cite{ST}) \( r=r_{i}\left(1-t/t_{0}\right)^{1/4},
\) where \( t_{0}= (5r_{i}^{4}c^{5})/(256\mu M_{t}^{2}G^{3}) \) and
$r_{i}$ is the initial separation at $t$=0. The total mass is
$M_{t}$=$M_{BH}$+$M$ and the reduced mass is
$\mu$=$M_{BH}M/M_{t}$. Clearly, hydrodynamical effects and the
instability they induce drive the coalescence process on a time scale
that is shorter than that purely due to gravitational radiation
losses, and so we may safely neglect the effect of radiation reaction
on the evolution of the system over the time period which we have
modeled.

\begin{figure}
\plotone{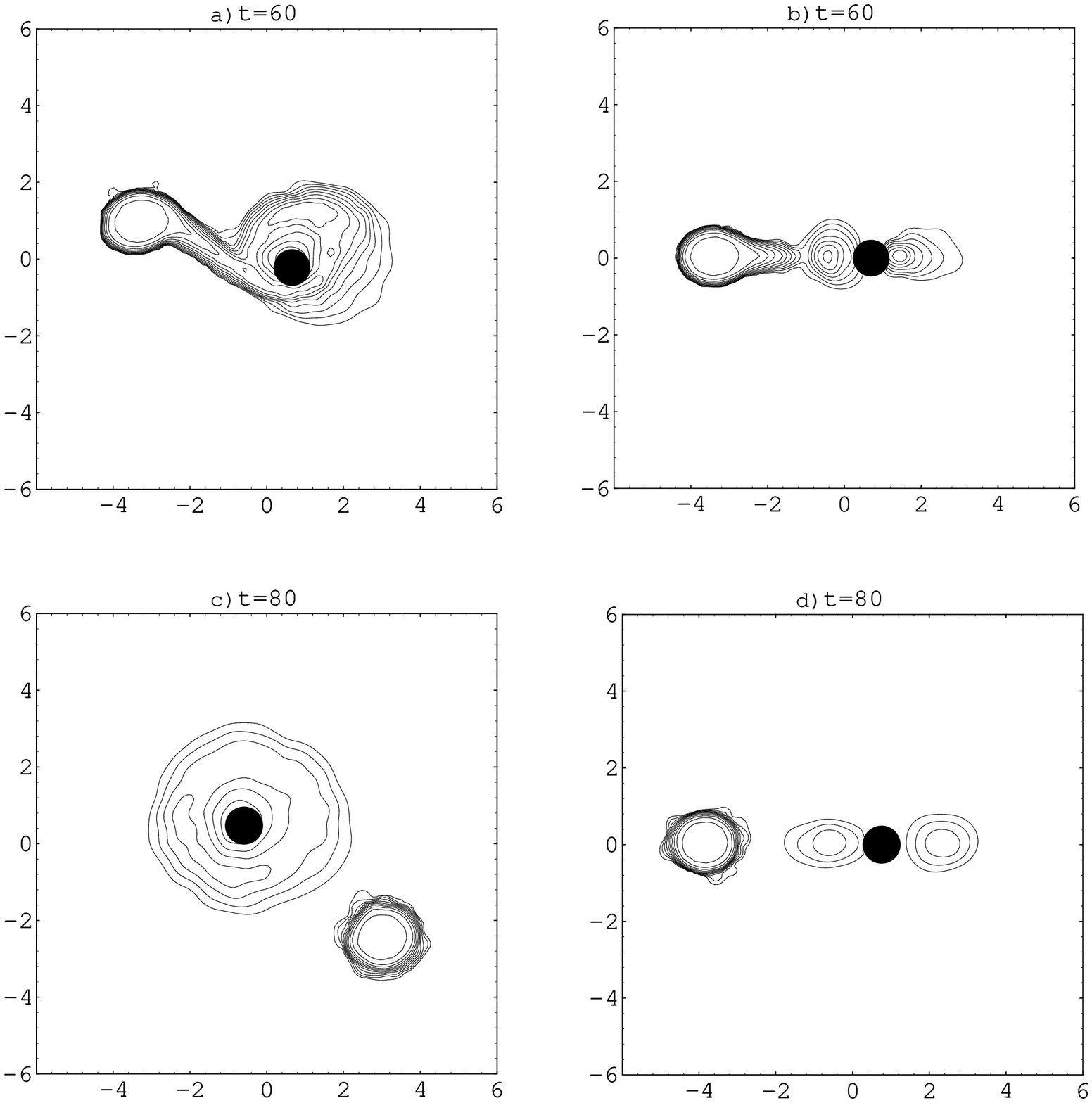}
\caption{Density contours for the dynamical
coalescence of the black hole--neutron star binary with mass ratio
$q$=1 and initial separation $r$=2.78, Case II. There are eleven
evenly spaced logarithmic contours down from $\log{\rho}=-0.75$ every
0.25 decades. The rotation is counterclockwise and the initial orbital
period is $P$=20.09. The left column shows density contours in the
orbital plane, the right column shows contours in the meridional plane
containing the binary axis. The black disk represents the black
hole. \label{rhoII}}
\end{figure}

\begin{figure}
\plotone{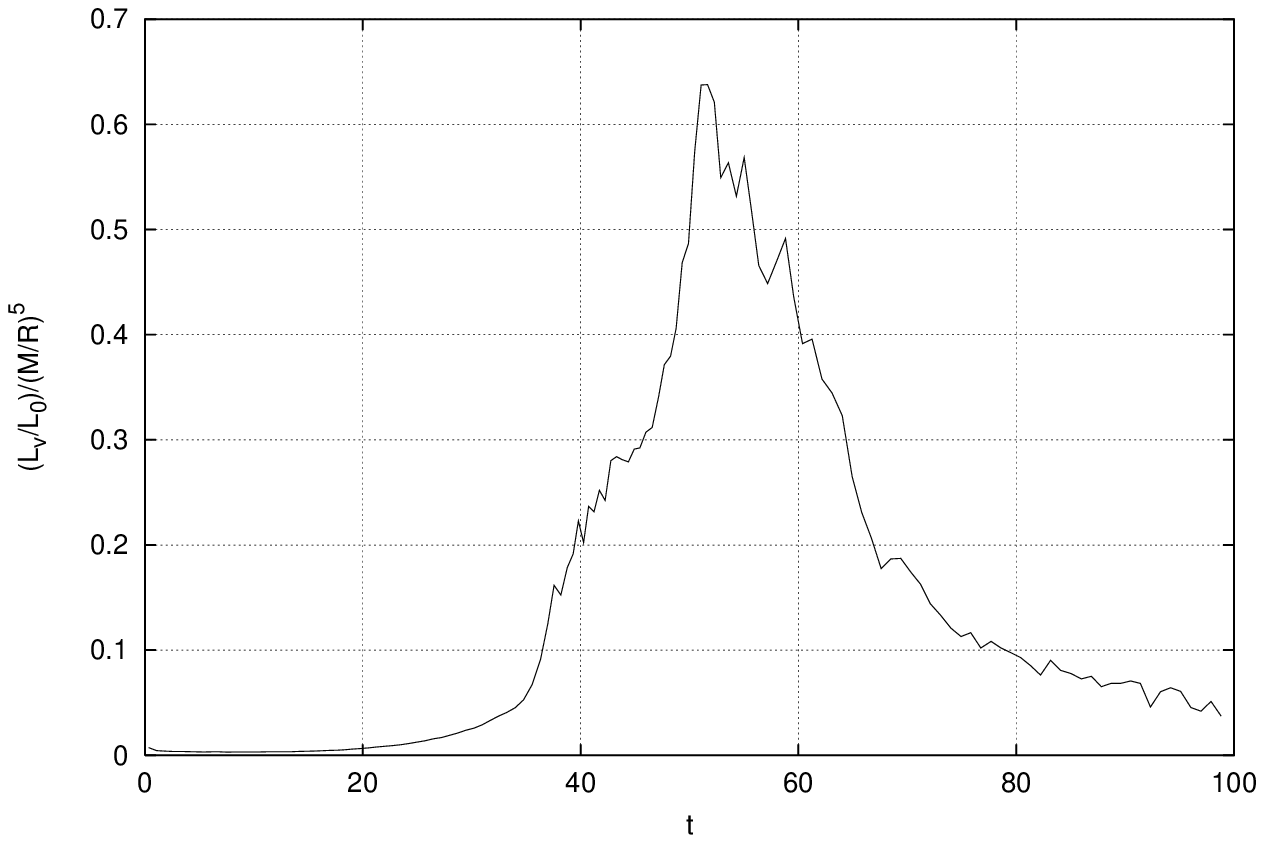}
\caption{The thermal energy loss rate due to viscosity $L_{v}$ in
geometrized units (such that $G$=$c$=1), normalized to
$L_{0}$=$c^{5}/G$=$3.6\times10^{59}$~erg~s$^{-1}$, as a function of
time for an initial configuration with a mass ratio $q$=1 and initial
separation $r$=2.78, Case II. \label{heatloss}}
\end{figure}

In the calculation described above, we have integrated the thermal
energy equation according to the first law of thermodynamics by
setting
\begin{eqnarray}
\Delta U=\Delta W+\Delta Q \label{eq:thermoI}
\end{eqnarray}
where $\Delta Q$ is calculated using the contribution from the
artifical viscosity (see equation~\ref{eq:dudt} in
Appendix~\ref{code}). We will denote this as Case I. To investigate
the effect of radiative losses on the outcome of this configuration,
we have performed an additional calculation for the extreme case in
which all the energy produced by viscosity is lost by the system
(e.g. to neutrinos). This will be denoted as Case II, and is achieved
by simply removing the term associated with viscous heating from
equation~(\ref{eq:thermoI}), so that we now have $\Delta U$=$\Delta
W$. Density contours for Case II (with $N$=16944 particles) are shown
in Figure~\ref{rhoII}
for $t$=60 and $t$=80. An accretion torus is again clearly visible,
and comparing with the contours in Figure~\ref{rhobhq1} for Case I,
the ring is clearly more confined both in the orbital plane and along
the $z$-axis. This is clearly a consequence of a lack of pressure
support due to the removal of thermal energy from the system. In
Figure~\ref{heatloss} 
we show the rate of energy loss, $L_{v}$ due to
viscous heating. The peak rate is $(R/M)^{5}L_{v}\approx 0.65$,
corresponding to $\approx 2\times 10^{55}$~erg~s$^{-1}$ and the total
energy loss from $t$=0 to $t$=100 is $\approx 5\times
10^{52}$~erg. The mass transfer rate and gravitational radiation
waveforms are essentially the same as for Case I. There is once more a
baryon--free axis along the rotation axis of the binary, only this
time it is clear of matter within $\sim 20^{\rm o}$ (down to $4\times
10^{-5}$, equivalent to $6\times 10^{-5}M_{\odot}$) since the torus is
more nearly confined to the orbital plane.

\subsection{Mass ratio $q$=0.8}

\subsubsection{Equilibrium configurations}

For a mass ratio $q$=0.8, the black hole mass is $M_{BH}$=1.25~(or
1.75$M_{\odot}$). Hydrodynamical effects are still important in the
construction of equilibrium configurations in this case, as can be
seen in 
\begin{figure}
\plotone{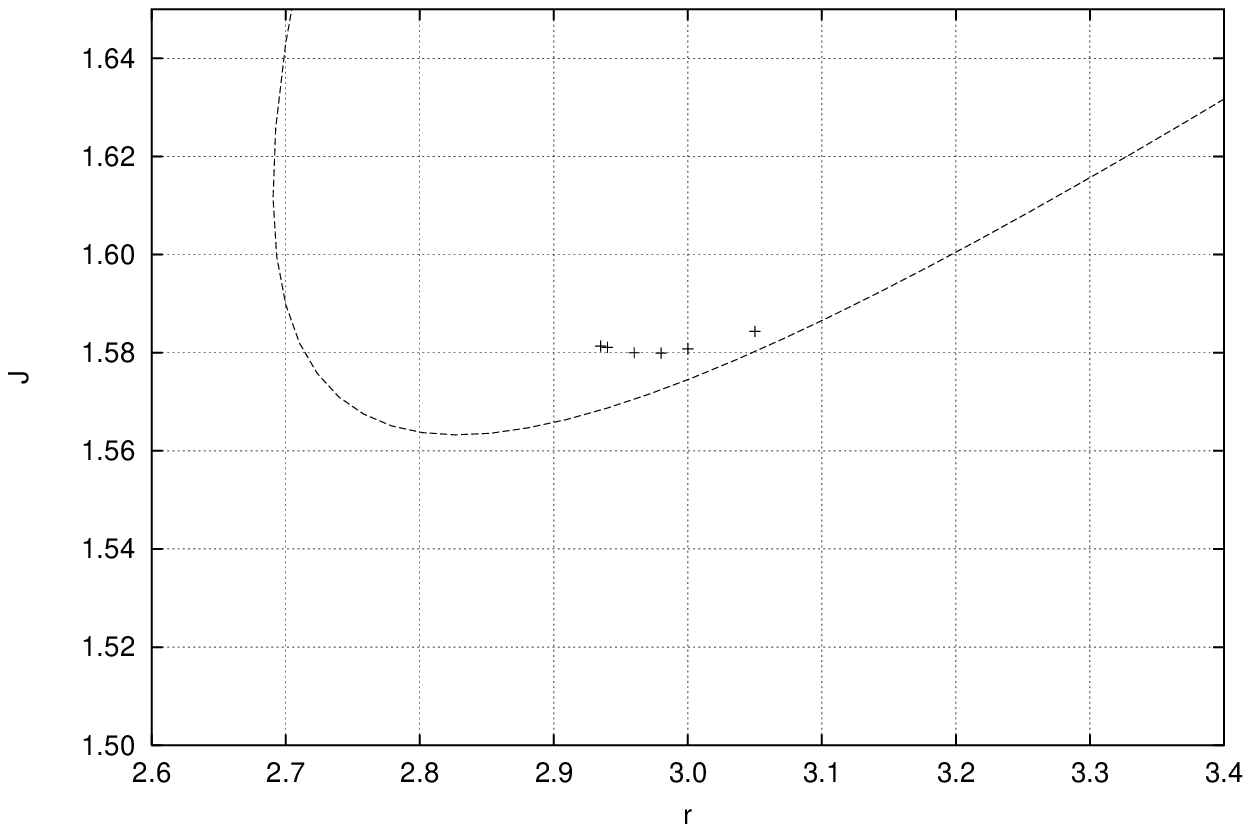}
\caption{Total angular momentum vs. binary separation for a black
hole--neutron star binary with mass ratio $q$=0.8. The solid line is
the result of the analytical approach of~\cite{LRSb} treating the
neutron star as a compressible tri--axial ellipsoid, and the crosses
are the result of SPH relaxation calculations. The Roche limit is at
$r$=2.935. \label{jvsrq08}}
\end{figure}
Figure~\ref{jvsrq08}, where we have plotted the equilibrium
angular momentum values for a range of binary
separations. Approximating the neutron star as a compressible
tri--axial ellipsoid gives results which are close to the full
numerical calculations before the minimum in angular momentum is
reached, but for $r\leq 3.0$ this is no longer the case. We identify
the separation with minimum angular momentum at $r$=2.98, and
Roche--Lobe overflow occurs at $r_{RL}$=2.935. For this value of the
mass ratio $N$=8121 particles model the neutron star.

\subsubsection{Dynamical runs} 

\begin{figure}
\plotone{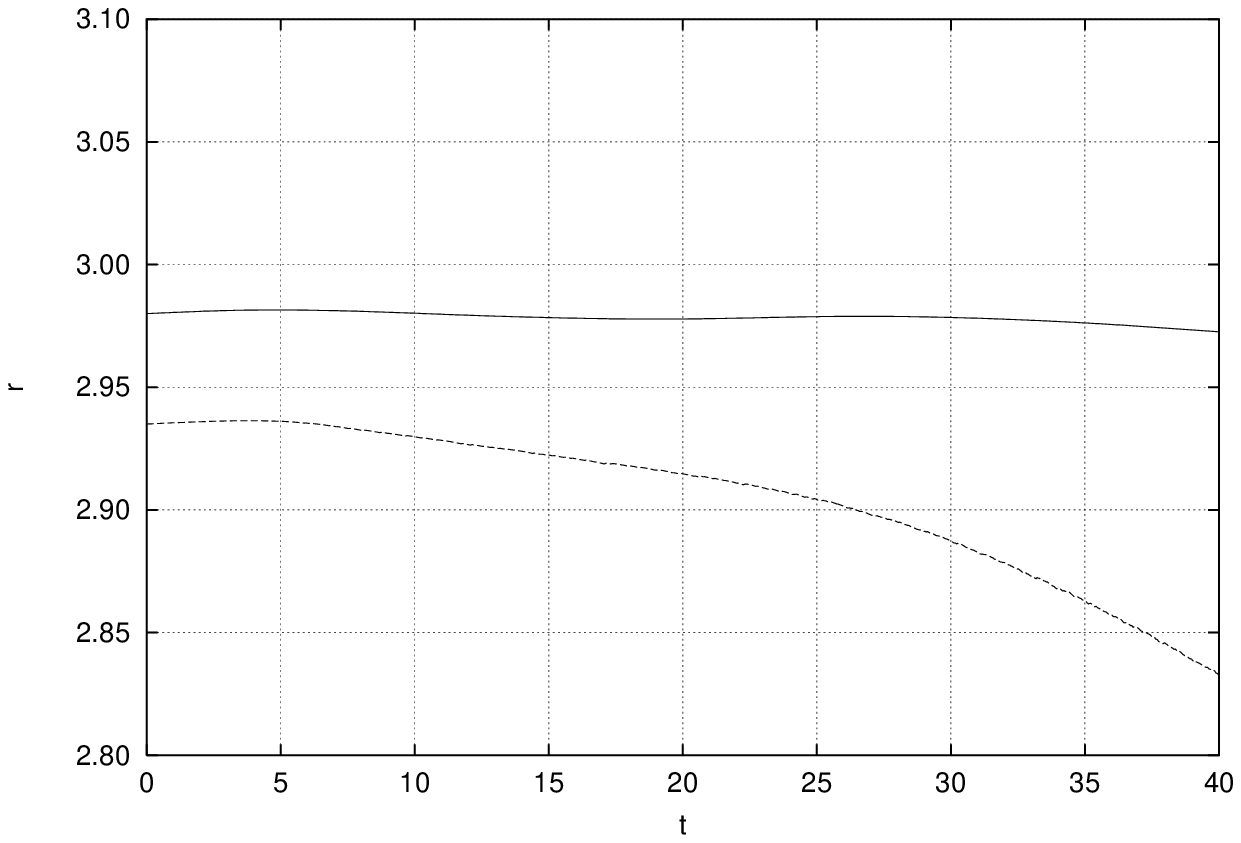}
\caption{Binary separation as a function of time for dynamical
calculations starting at $r$=2.98 (solid line) and $r$=2.935 (dotted
line) for the black hole--neutron star binary with mass ratio
$q$=0.8. \label{rvstq08}}
\end{figure}

As for the case with $q$=1, several equilibrium configurations were
used as initial conditions for dynamical runs. Figure~\ref{rvstq08}
shows 
the binary separation as a function of time for these
calculations. The initial configuration with separation $r$=2.935
becomes unstable on a dynamical time scale. However, it is apparent
that the development of the instability (and hence the orbital decay)
proceeds at a slower rate than for a higher mass ratio. The reason for
this is that since the black hole is more massive, Roche--Lobe
overflow occurs at a greater separation than previously observed (for
$q$=1), and so the tidal effects are less pronounced on the neutron
star. Density contours in the orbital plane for the binary with
initial separation $r$=2.935 are 
\begin{figure}
\plotone{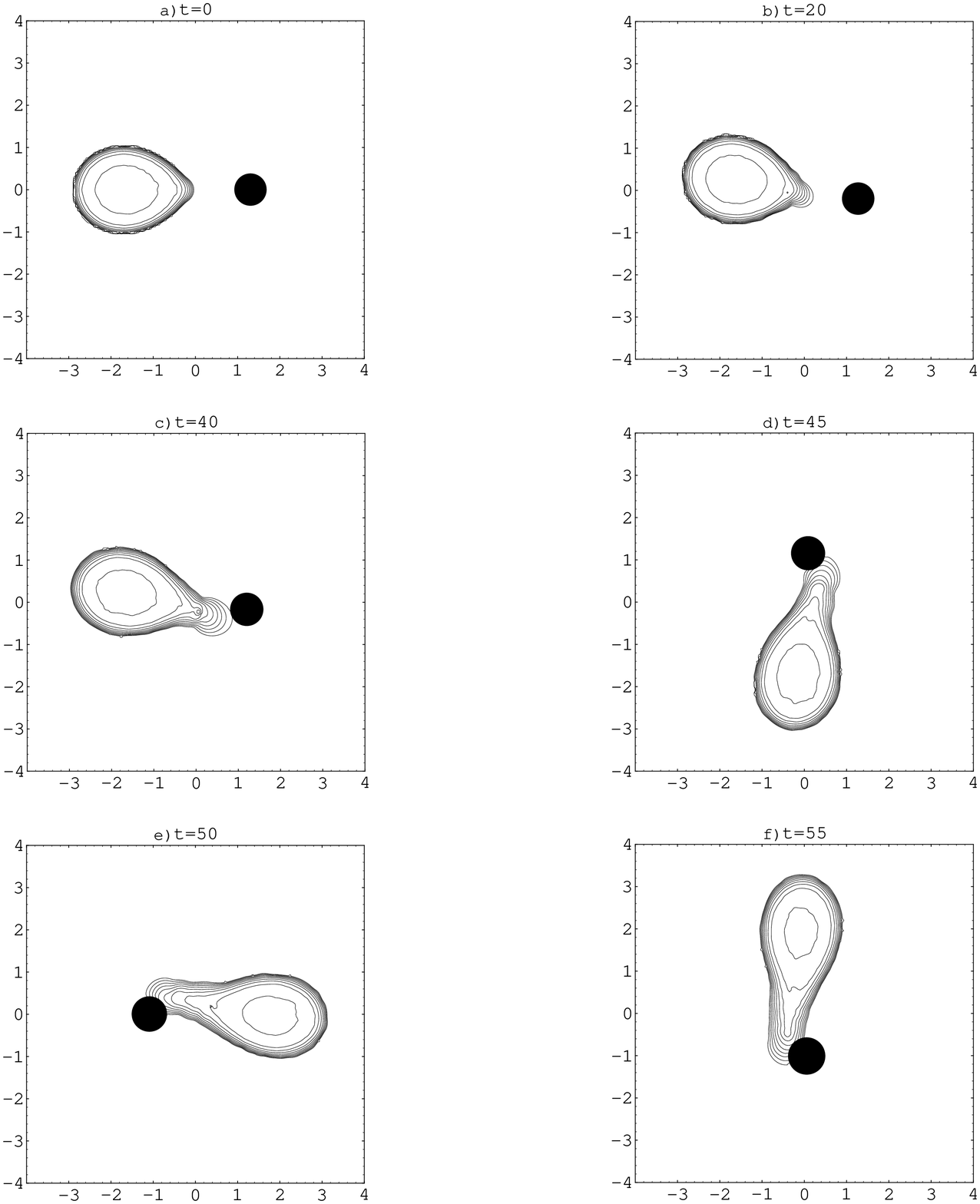}
\caption{Density contours in the orbital plane for
a black hole--neutron star coalescence with a mass ratio $q$=0.8 and
initial separation $r$=2.935. The logarithmic contours are evenly
spaced every 0.25 decades, with the lowest one at
$\log{\rho}=-2.75$. The black disk represents the black hole. Orbital
rotation is counterclockwise and the initial orbital period is
$P$=20.58. \label{bhnsq08contours}}
\end{figure}
\setcounter{figure}{11}
\begin{figure}
\plotone{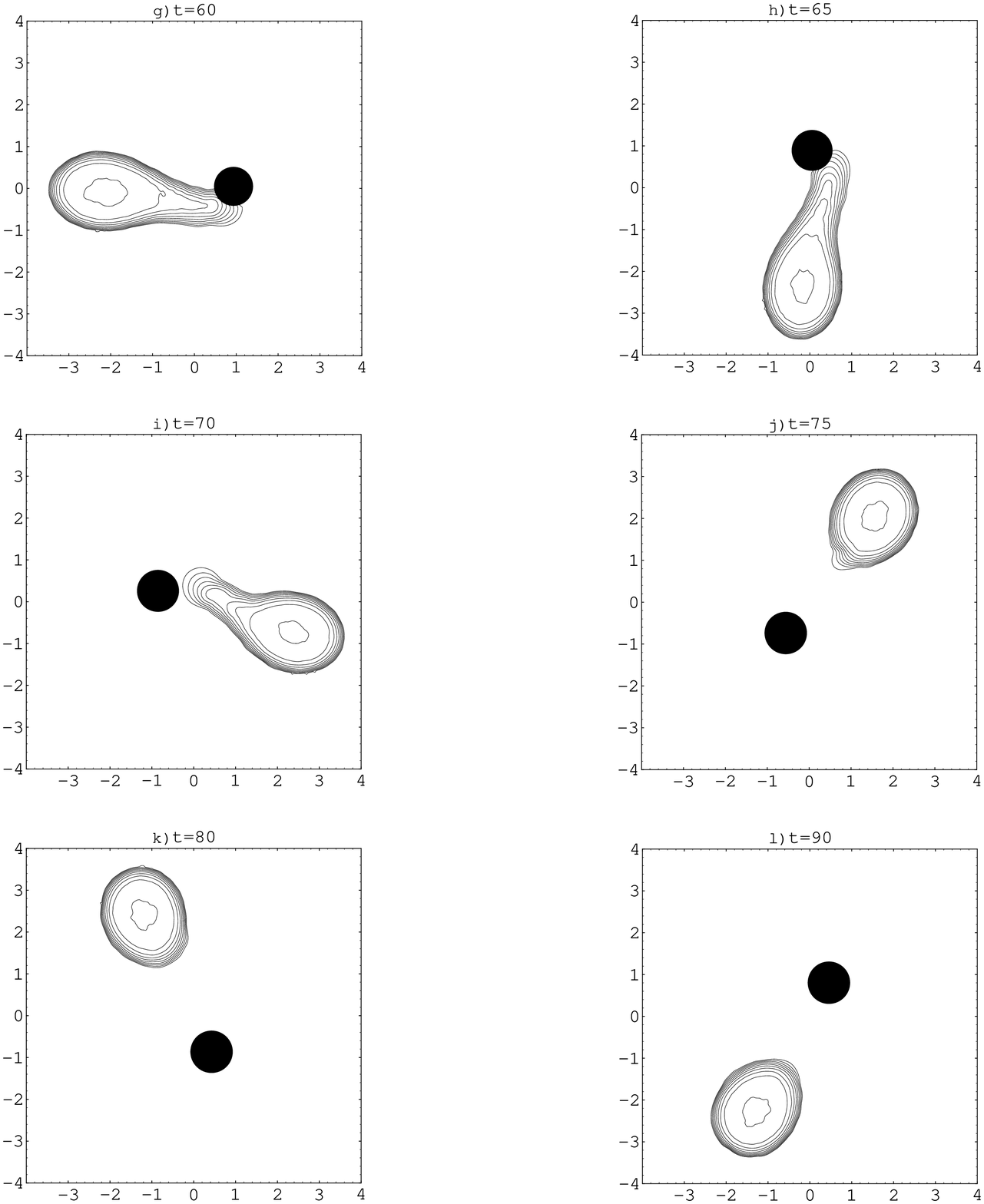}
\caption{continued.}
\end{figure}
shown in Figure~\ref{bhnsq08contours}. Mass transfer proceeds through
the formation of an accretion stream from the neutron star to the
black hole, lasting from $t$$\sim$45 to $t$$\sim$75 (about 3.5~ms),
with a peak accretion rate of 0.017~(0.2$M_{\odot}$/ms). A substantial
amount of mass is stripped from the neutron star during the encounter
(see Figure~\ref{mbhq08}), and
\begin{figure}
\plotone{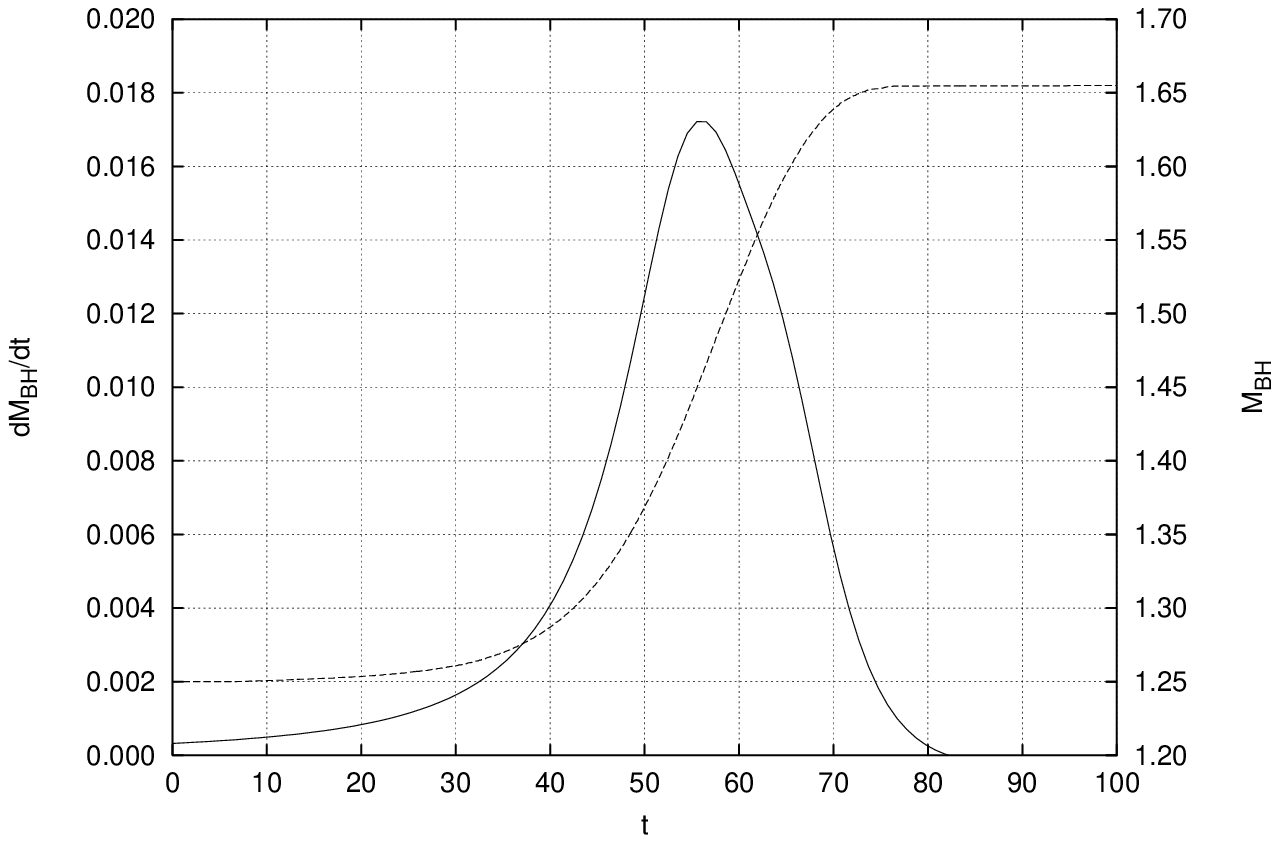}
\caption{Black hole mass (dotted line) and mass accretion rate onto
the black hole (solid line) as a function of time for the black
hole--neutron star binary with mass ratio $q$=0.8 and initial
separation $r$=2.935. \label{mbhq08}}
\end{figure}
the final black hole mass is
$M_{BH}$=1.65~(equivalent to 2.3$M_{\odot}$). The core of the neutron
star, with a final mass $M_{core}$=0.595~(corresponding to
0.83$M_{\odot}$) again survives the encounter. However, the most
striking difference between this case and the one presented previously
($q$=1) is the absence of a toroidal accretion structure around the
black hole. All matter stripped from the neutron star is directly
accreted by the black hole, and thus in this scenario as well the
rotation axis above the black hole is devoid of matter. As a result of
mass transfer the binary separation increases to $r$$\sim$3.6 and thus
the final configuration in this case is again a stable binary.

\begin{figure}
\plotone{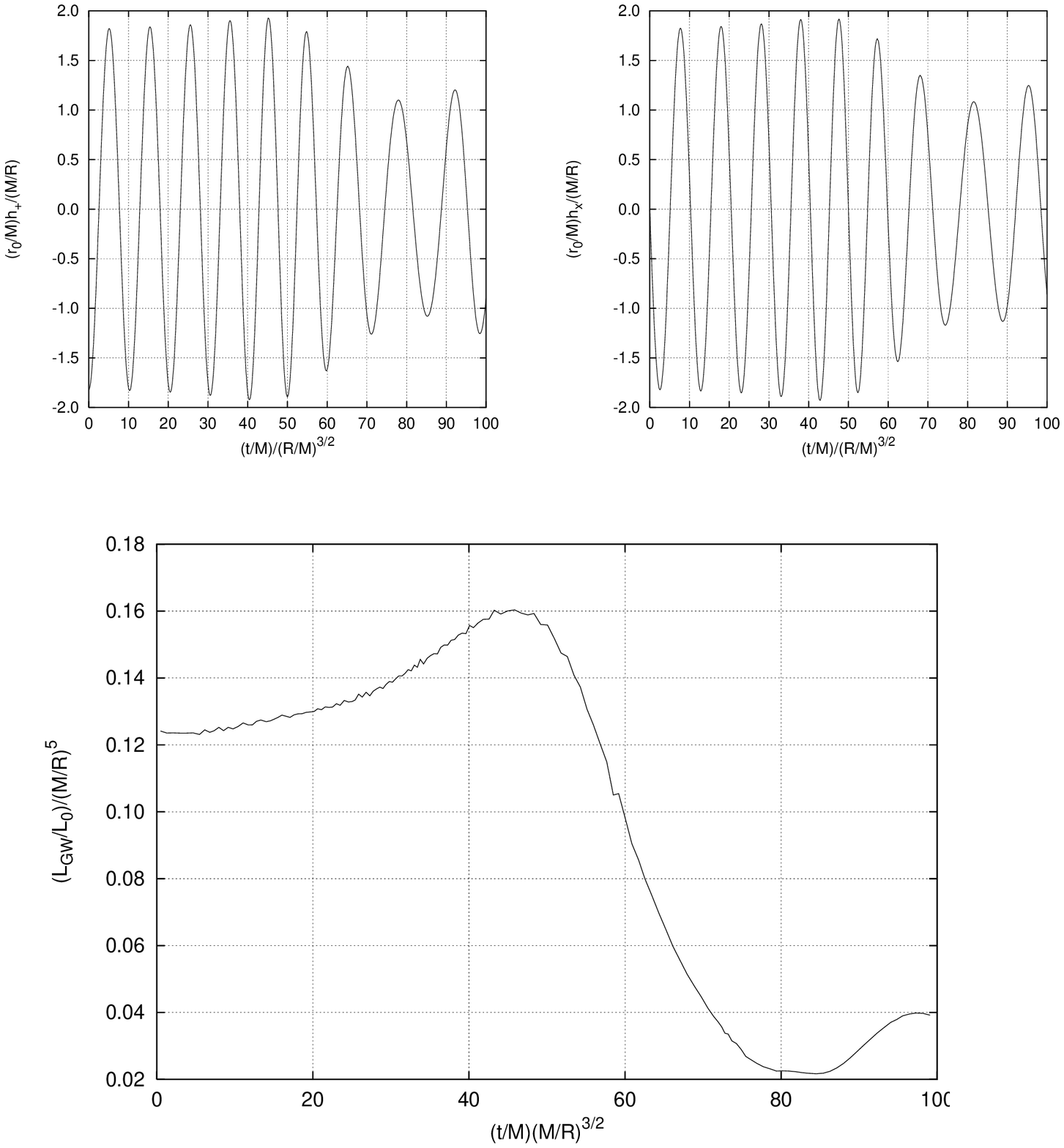}
\caption{Top: Gravitational radiation waveforms for the coalescence of
a black hole--neutron star binary with mass ratio $q$=0.8. Two
polarizations ($h_{+}$, left panel; $h_{\times}$, right panel) in
geometrized units (such that $G$=$c$=1) are shown. Bottom:
Gravitational radiation luminosity in geometrized units
($G$=$c$=1) for the coalescence of a black hole--neutron star binary
with mass ratio $q$=0.8. The constant
$L_{0}$=$c^{5}/G$=$3.6\times10^{59}$~erg~s$^{-1}$. \label{bhnsq08grav}}
\end{figure}

As before, the hydrodynamical evolution of the binary is reflected in
the gravitational radiation waveforms, presented in
Figure~\ref{bhnsq08grav}~(top). 
There is an initial rise in amplitude in the waveforms, followed by a
rapid decline and increase in the period, reflecting the episode of
mass transfer and subsequent increase in binary separation. The peak
amplitude in this case is $(r_{0}R/M^{2})h_{max}$=1.93, and again
since the binary survives the encounter, the final amplitude is not
zero but $(r_{0}R/M^{2})h_{final}$=1.25. The luminosity emitted in
gravitational waves is shown in Figure~\ref{bhnsq08grav}~(bottom).
The general shape of the curve resembles very much that for $q$=1, as
one can expect after studying the density contour plots for each case
(since the toroidal structure that was formed for $q$=1 is essentially
azimuthally symmetric it does not contribute significantly to the
emission of gravitational waves). The curve presents again a broad
peak, with a duration $\Delta t$$\sim$30~(3.4~ms). The peak luminosity
is $(R/M)^{5}L_{max}$=0.16~(or $4.8\times 10^{54}$~erg~s$^{-1}$) and
the total energy radiated away in gravitational waves is
$(R^{7/2}/M^{9/2})\Delta E_{GW}$$\approx$$10$~(or $3.4\times
10^{52}$~erg). These values are higher than for the case with $q$=1
because we have normalized our results to the mass of neutron star,
and lowering the mass ratio implies the system has a greater total
mass.

\begin{figure}
\plotone{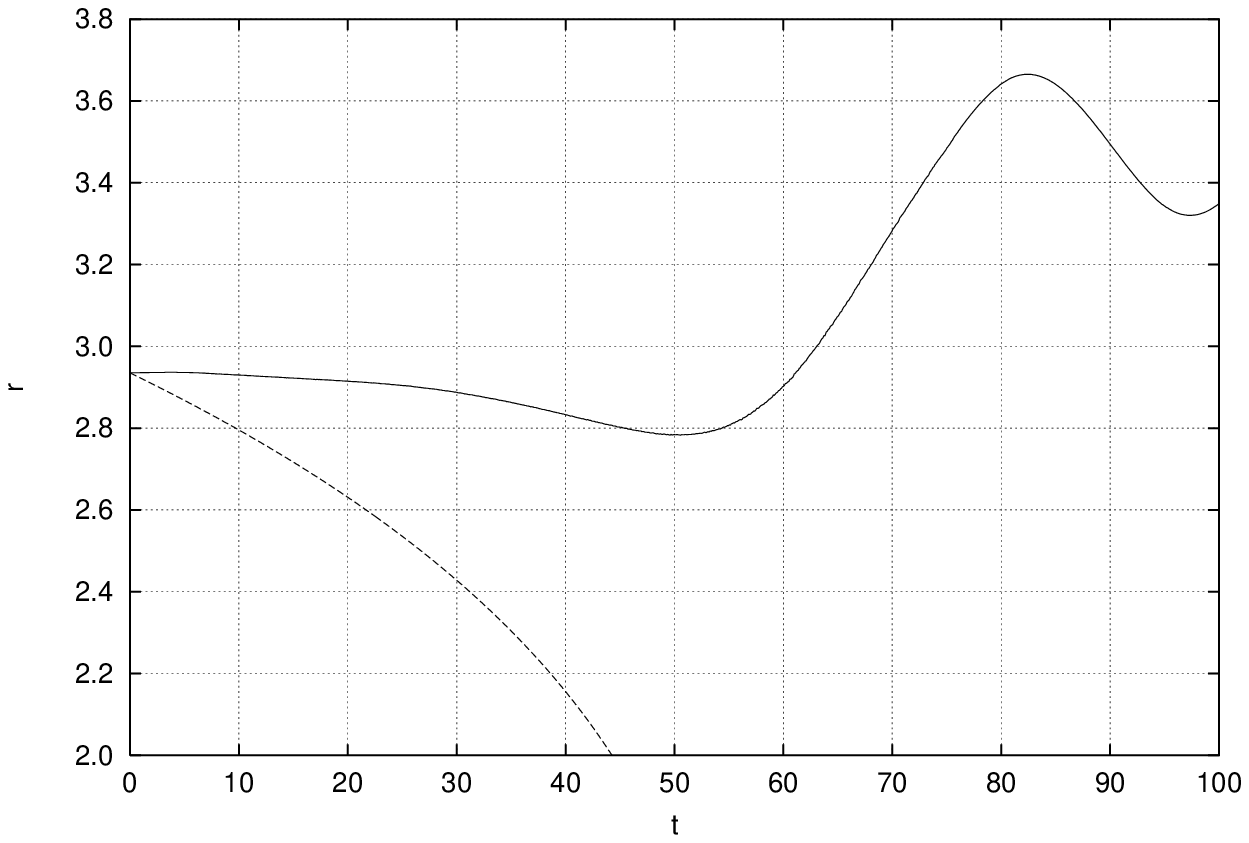}
\caption{Binary separation for the coalescence of a black
hole--neutron star binary with mass ratio $q$=0.8 (solid line) and for
a point--mass binary with the same mass ratio and initial separation
decaying via the emission of gravitational waves in the quadrupole
approximation (dotted line). \label{ptdecayq08}}
\end{figure}

In Figure~\ref{ptdecayq08} 
we compare (as for $q$=1 in Figure~\ref{ptdecayq1}) the orbital decay
of this binary with that for a point--mass binary emitting
gravitational waves in the quadrupole approximation. It is apparent
that hydrodynamical effects are less important than for $q$=1 in
determining the initial orbital evolution, but are nevertheless
substantial.

\subsection{Mass ratio $q$=0.31} 

\begin{figure}
\plotone{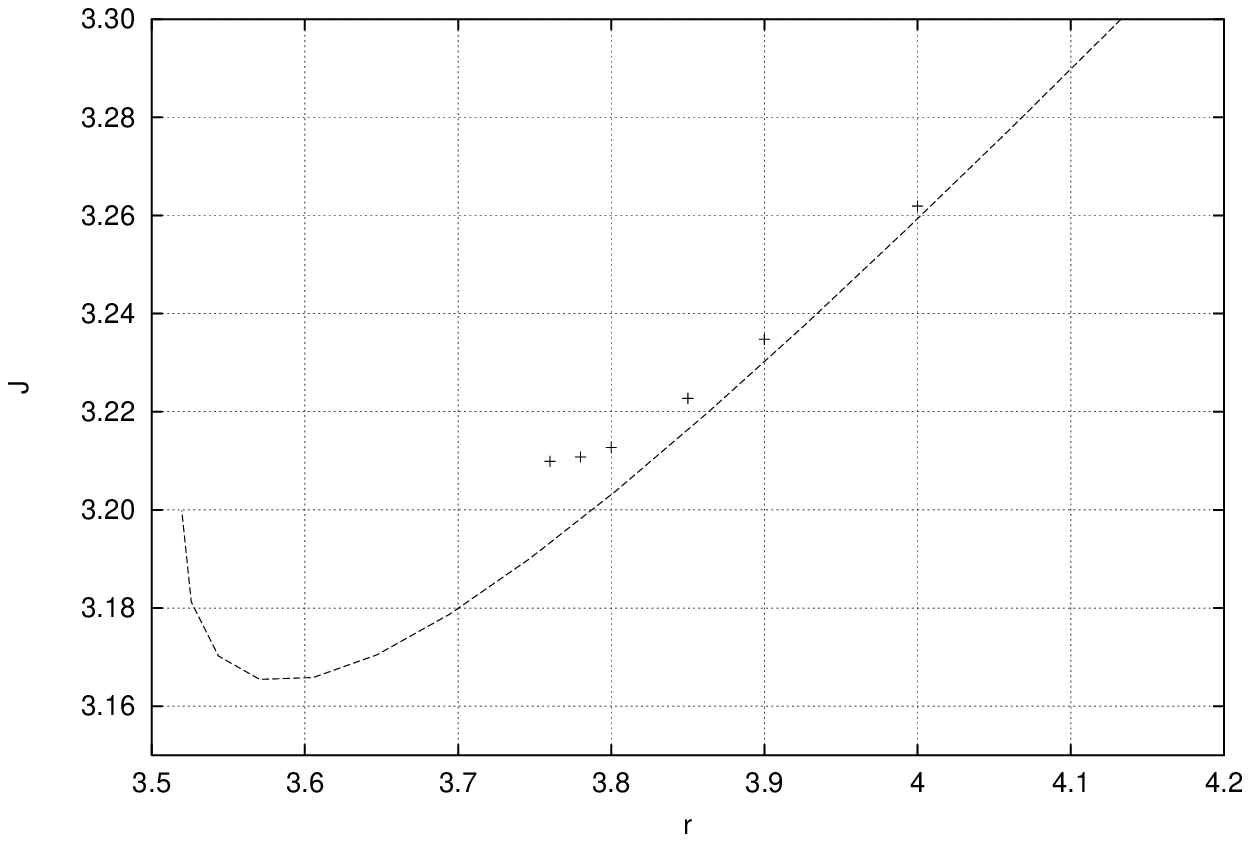}
\caption{Total angular momentum vs. binary separation for a black
hole--neutron star binary with mass ratio $q$=0.31. The solid line is
the result of the analytical approach of~\cite{LRSb} treating the
neutron star as a compressible tri--axial ellipsoid, and the crosses
are the result of SPH relaxation calculations. The Roche limit is at
$r$=3.76. Note that Roche--Lobe overflow is reached {\em before} the
minimum in angular momentum. \label{jvsrq031}}
\end{figure}

Continuing the trend towards lower mass ratios, we chose a value of
$q$=0.31, giving $M_{BH}$=3.22~(corresponding to 4.51$M_{\odot}$) to
study the evolution of a binary system where most of the mass is
contained in the black hole. We again used $N$=8121 particles to model
the neutron star. A qualitative difference appears in the plot of
total angular momentum as a function of binary separation
(Figure~\ref{jvsrq031}). 
At a separation of $r$=3.76 (or 50.5~km),
Roche--Lobe overflow occurs, so that we cannot construct a sequence
with decreasing separation and constant mass ratio anymore, but this
point is reached {\em before} the minimum in total angular momentum is
attained, contrary to what occurred for $q$=1 and $q$=0.8. Thus we
expect all configurations down to the Roche limit $r_{RL}$=3.76 to be
dynamically stable (see the discussion for $q$=1 in
Section~\ref{initq1}).

\begin{figure}
\plotone{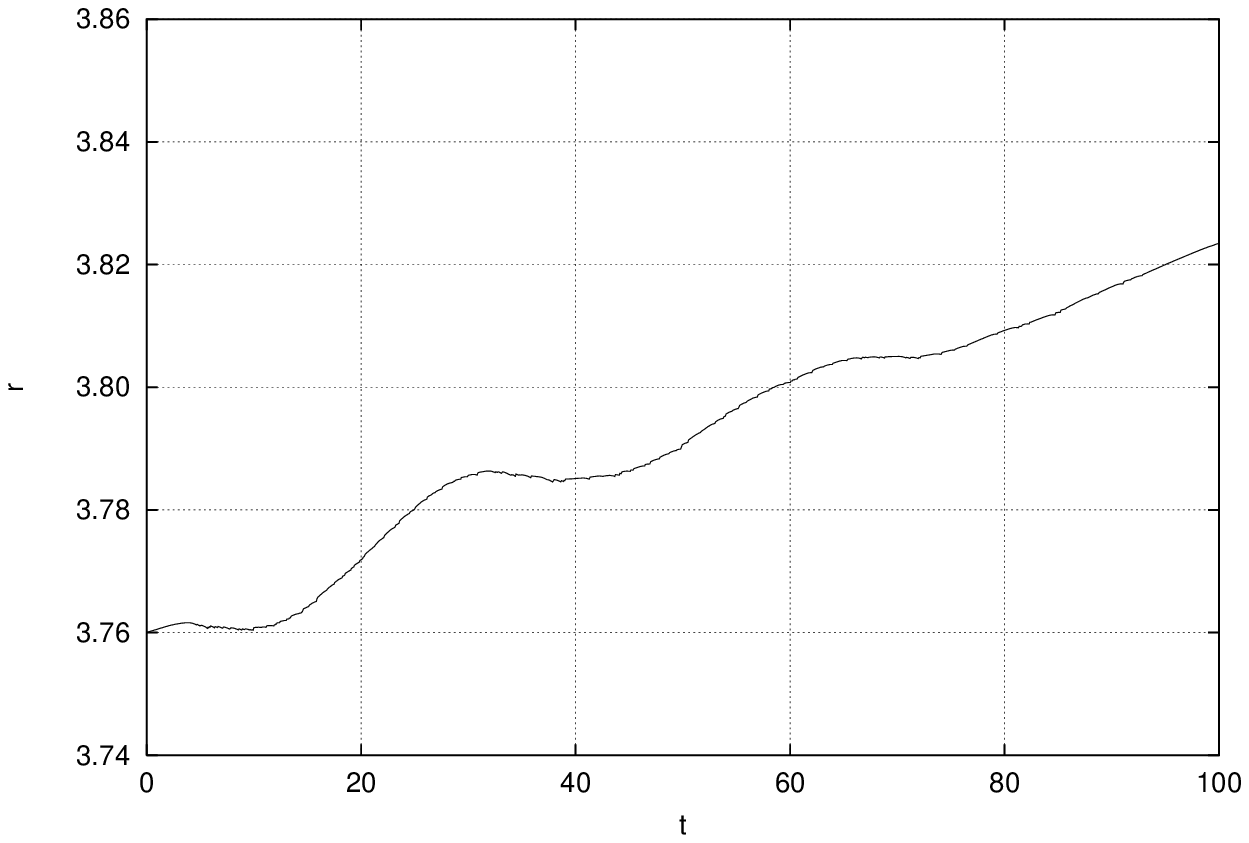}
\caption{Binary separation as a function of time for a dynamical
calculation starting at $r$=3.76 for the black hole--neutron star
binary with mass ratio $q$=0.31. \label{rvstq031}}
\end{figure}

\begin{figure}
\plotone{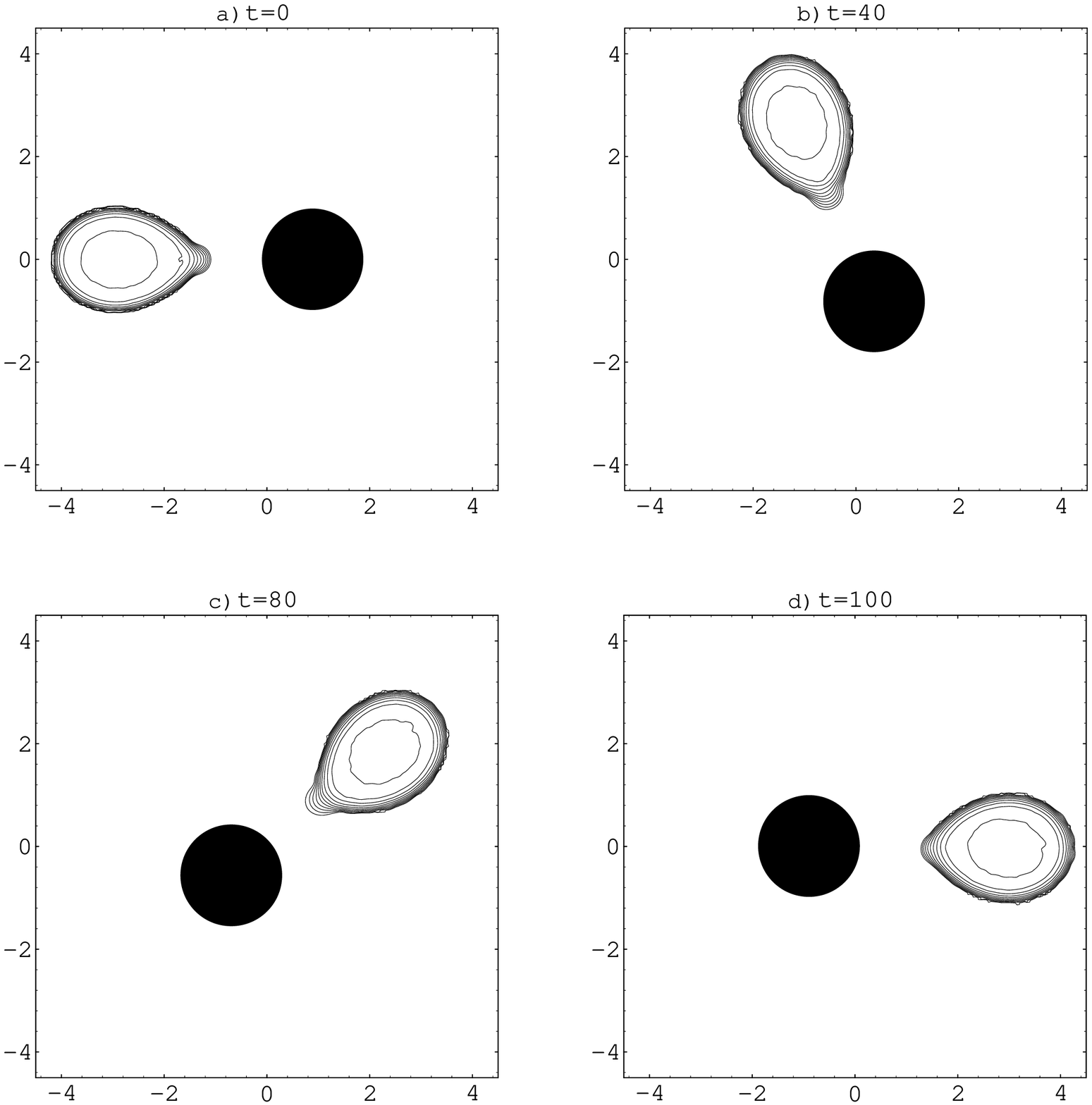}
\caption{Density contours at various times during
the dynamical coalescence of the black hole--neutron star binary with
mass ratio $q$=0.31 and initial separation $r$=3.76. The logarithmic
contours are evenly spaced every 0.25 decades, with the lowest one at
$\log{\rho}$=-3.0. The rotation is counterclockwise and the initial
orbital period is $P$=21.92. The black disk represents the black
hole. \label{rhobhq031}}
\end{figure}

\begin{figure}
\plotone{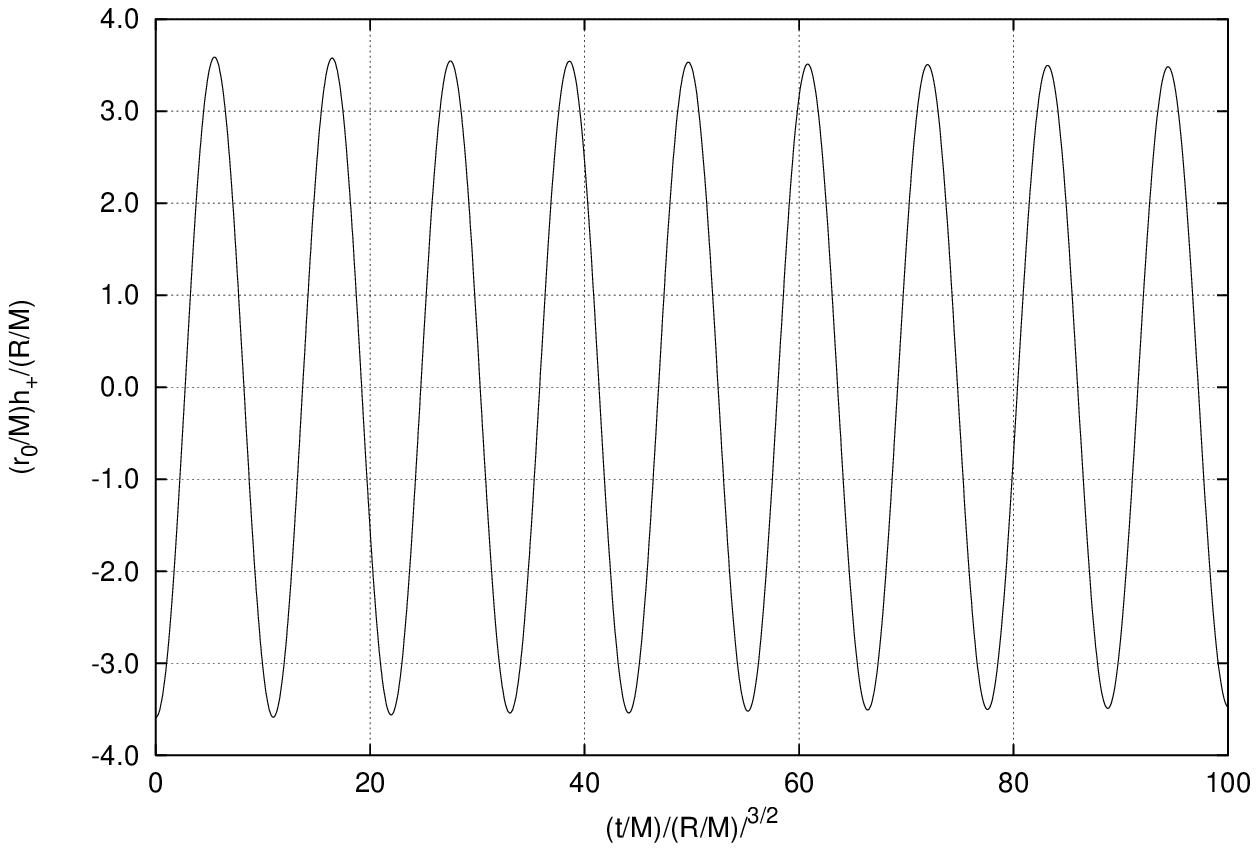}
\caption{Gravitational radiation waveforms for the
coalescence of a black hole--neutron star binary with mass ratio
$q$=0.31. One polarization ($h_{+}$) in geometrized units (such that
$G$=$c$=1) is shown. \label{bhnsq031h}}
\end{figure}

The binary separation for a dynamical run with an initial separation
$r$=$r_{RL}$=3.76 is shown in Figure~\ref{rvstq031}. 
The initial
orbital period is $P$=21.92~(or 2.51~ms). There is no longer orbital
decay, but rather a slow increase in separation. This is simply due to
the fact that there is slow Roche--Lobe overflow from the neutron star
onto the black hole, transferring $\Delta M$=0.0104~(or
0.0146$M_{\odot}$) over $\sim$4.5 orbital periods. The neutron star is
not disrupted at all during this process, as can be seen in
Figure~\ref{rhobhq031}, 
where we plot density contours in the orbital
plane over the course of the simulation, and in
Figure~\ref{bhnsq031h}, 
where we plot the gravitational radiation
waveform $h_{+}$ for this case. In fact, mass transfer is essentially
conservative (i.e. the total angular momentum is conserved), and this
approximation alone can accurately account for the change in
separation. To see this, consider the following. Conservative mass
transfer in a point--mass binary satisfies \( J=\mu \Omega
a^{2}=constant\), where {\em a} is the separation and $\mu$ is the
reduced mass. Since the total mass is conserved and $\Omega\propto
a^{-3/2}$ for a Keplerian binary, we find that \( (M_{BH}M)^{2}
a=constant \). For the black hole--neutron star binary with initial
mass ratio $q$=0.31, we find that
\begin{eqnarray*}
\left[ \frac{(M_{BH}M)_{initial}}{(M_{BH}M)_{final}}
\right]^{2}=1.01453
\end{eqnarray*}
and
\begin{eqnarray*}
\frac{a_{final}}{a_{initial}}=1.0167.
\end{eqnarray*}
This is consistent with having conservative mass transfer despite the
fact that the black hole--neutron star binary is not
Keplerian. Angular momentum is practically constant since there is
very little accretion, and this represents the only loss of angular
momentum (via a spinning up of the black hole, which we do not model).

This last case is a clear indication of the limitations of the present
method of studying the coalescence of a black hole with a neutron
star. In the first place, since the orbit does not decay due to
hydrodynamical effects, gravitational radiation reaction (which would
drive the binary mentioned above to coalesce in a time $t_{0}$=31.35,
correponding to 3.59~ms) cannot be ignored. Second, as is apparent in
the density contour plots shown in Figure~\ref{rhobhq031}, the black
hole has become almost as large as the neutron star itself (the
Schwarschild radius is $r_{s}$=0.986, or 13.2~km, at the start of the
simulation), making the radius of the innermost circular stable
orbit~(ISCO) $r_{ISCO}$=$3r_{s}$=2.958 (or 39.75~km). This means a
substantial fraction of the neutron star is already at a distance
$r\leq r_{ISCO}$ from the black hole. Thus including relativistic
effects in this case becomes necessary in order to obtain meaningful
results.

\section{Discussion \label{discussion}}

We present in Table~\ref{critical} the binary separations $r_{RL}$ and
$r_{min}$, corresponding to the Roche limit and to the equilibrium
configuration with minimum total angular momentum $J$ for each mass
ratio. The lack of an entry for $r_{min}$ when $q$=0.31 reflects the
fact that the orbits are stable when the separation is greater than
that required for Roche--Lobe overflow.

In Table~\ref{summary} we present a summary for the three
configurations we explored in this paper, as well as for the
coalescence of two identical neutron stars. The first column indicates
if the coalescence involves two neutron stars or one neutron star and
one black hole; the second and third columns show the initial and
final mass ratios respectively; the fourth column indicates whether an
accretion torus was formed; the fifth, sixth and seventh columns show
the maximum and final gravitational radiation amplitude and the
maximum gravitational radiation luminosity respectively, all in
geometrized units such that $G$=$c$=1; the eighth column shows the
mass of the central object for the case of the double neutron star
coalescence, and that of the remnant core in the case of the black
hole--neutron star coalescence.

Note that only in the simulation with a high mass ratio ($q$=1) did an
accretion structure appear around the black hole.  Otherwise, all
matter stripped from the neutron star was directly accreted by the
black hole. A striking result is that in every case, the neutron star
is not completely disrupted, but a remnant core survives in orbit
around the more massive black hole. Due to the violent mass transfer
episode (for $q$=0.8 and $q$=1), this core is transferred to a higher
orbit on a time scale of one orbit. For now, we are unable to model
the system on longer time scales than we have presented here, but
presumably the emission of gravitational waves will make the orbit
decay by decreasing the separation. Clearly mass transfer will occur
before the separation reaches zero, as soon as the neutron star
overflows its Roche Lobe. For the equation of state we have considered
(with $\Gamma$=3), the mass--radius relationship is $R\propto M^{1/5}$
and one can estimate the size of the Roche Lobe as
$R_{RL}$=0.46$a(M/M_{t})^{1/3}$~(\cite{bp67}). Using the rate of
angular momenutum loss calculated from the quadrupole approximation,
given by~(\cite{ST})
\begin{eqnarray*}
\frac{dJ}{dt}=-\frac{32}{5}\frac{G^{3}M_{t}^{2}\mu}{c^{5}a^{4}},
\end{eqnarray*}
we can find the separation at which mass transfer will begin anew, as
well as the time it will take for the orbit to shrink to this
size. Applying this reasoning to the binary with initial $q$=1 after
the initial mass transfer event, we find that a second episode of mass
transfer will occur $\sim$40~ms later, and for the case with initial
$q$=0.8, the delay will be $\sim$4~ms. Thus the length of the
coalescence process is extended from a few milliseconds to possibly
several tens of milliseconds.

Another important result is the fact that in every case, there is a
baryon--free line of sight to the black hole along the rotation axis
of the binary throughout the simulation. As stated in
section~\ref{dynq1}, the limit we infer for the amount of matter along
this axis from our calculations is $10^{-4}M_{\odot}$ within about
5$^{\rm o}$ of the rotation axis. It is set by the resolution in our
calculation and thus represents an upper bound at this point. For the
calculations with a lower mass ratio ($q$=0.8 and $q$=0.31) there was
no visible accretion structure present around the black hole, and so
the rotation axis was clear of matter as well. These last two results
may make the coalescence of a black hole with a neutron star a
promising candidate source for the production of GRBs~(\cite{KL}).

The survival of the neutron star core suggests a different outcome
than what was initially expected for the coalescence of a black hole
with a neutron star~(\cite{wheel}), and identifies this process as the
only one known capable of producing low--mass neutron stars. The
violent episode of mass transfer strips the neutron star of so much
matter that it may drive it below the minimum mass required for
stability. If this happens, then the core
explodes~(\cite{blin,sumiyoshi,colpi}) in approximately 0.1~seconds after a
slow expansion phase that may last 20~seconds~(\cite{sumiyoshi}). The
presence of the black hole would complicate the outcome of this event.

\acknowledgments

This work was supported in part through KBN grant 2P03D01311 and
DGDPA--UNAM.

\appendix

\section{Numerical Code \label{code}}

The equations of motion in SPH are essentially those of an N--body
problem, with each particle representing a fluid element. Thus the
equations for particle {\em i} can be written as:
\begin{eqnarray*}
\dot{\mbox{\boldmath$r$}}_{i} = \mbox{\boldmath $v_{i}$}, \label{eq:motion1}
\end{eqnarray*}
\begin{eqnarray*}
m_{i} \dot{\mbox{\boldmath $v$}}_{i} =\mbox{\boldmath
$F_{iG}$+$F_{iH}$}, \label{eq:motion2}
\end{eqnarray*}
where \mbox{\boldmath $r_{i}$}, \mbox{\boldmath $v_{i}$} and {\em
m}$_{i}$ are the position, velocity and mass of particle {\em i}
respectively, and \mbox{\boldmath $F_{iG}$} and \mbox{\boldmath
$F_{iH}$} denote the gravitational and hydrodynamic forces. The
density at the position of particle {\em i} is
\begin{eqnarray*}  
\rho_{i} \equiv \rho(\mbox{\boldmath$r_{i}$})= \sum_{j} m_{j} W_{ij}, 
\end{eqnarray*}
where $W_{ij}$ is the symmetrized kernel for particles {\em i} and
{\em j}. With $h_{i}$ representing the smoothing length for particle
{\em i}, we have
\begin{eqnarray*}
W_{ij}=W(\mid \mbox{\boldmath $r_{i}-r_{j}$} \mid,h_{ij}),\
h_{ij}=\frac{1}{2}(h_{i}+h_{j}). \label{eq:defWij}
\end{eqnarray*}

It is convenient to perform calculations with a kernel that has
compact support. We use the form of~\cite{ML}:
\begin{eqnarray*}
W(r,h) = \frac{1}{\pi h^{3}}\left\{ \begin{array}{ll} 1-\frac{3}{2}
		    \left( \frac{r}{h} \right)^{2}+\frac{3}{4} \left(
		    \frac{r}{h} \right) ^{3}, & 0 \leq r/h < 1, \\
		    \frac{1}{4} \left(2-\frac{r}{h} \right) ^{3}, & 1
		    \leq r/h < 2, \\ 0, & 2 \leq (r/h).  \end{array}
		    \right. \label{eq:defW}
\end{eqnarray*}

To obtain a value of the pressure for any particle, we use the
equation of state. For an ideal gas this becomes
\begin{eqnarray*}
P_{i} = (\Gamma-1) u_{i} \rho_{i}, \label{eq:eos}
\end{eqnarray*}
where $\Gamma$ is the adiabatic index and $u_{i}$ is the thermal
energy per unit mass for particle {\em i}.

\mbox {\boldmath $F_{iH}$} includes the contribution from the pressure
gradient and from artificial viscosity. In symmetrized form, it can be
written as
\begin{eqnarray}
\mbox{\boldmath $F_{iH}$}= - \sum_{j} m_{j}m_{i} \left(
2\frac{\sqrt{P_{i}P_{j}}}{\rho_{i}\rho_{j}} +\Pi_{ij} \right)
\mbox{\boldmath $\nabla$}_{i}W_{ij} , \label{eq:defhydrof}
\end{eqnarray}
where
\begin{eqnarray*}
\Pi_{ij} = \left\{ \begin{array}{ll} (-\alpha \overline{c}_{ij}
                   \mu_{ij} + \beta
                   \mu_{ij}^{2})/\overline{\rho}_{ij}, &
                   \mbox{\boldmath $v_{ij}$} \cdot \mbox{\boldmath
                   $r_{ij}$} <0 \\ 0, & \mbox{\boldmath $v_{ij}$}
                   \cdot \mbox{\boldmath $r_{ij}$} >0 \end{array}
                   \right. ,\label{eq:defviscosity}
\end{eqnarray*}
\begin{eqnarray*}
\mu_{ij}= \frac{h_{ij}(\mbox{\boldmath $v_{ij}$} \cdot \mbox{\boldmath
$r_{ij}$} )}{r_{ij}^{2} + \eta^{2} h_{ij}^{2}}. \label{eq:defmu}
\end{eqnarray*}
Here \mbox{\boldmath $v_{ij}$}=\mbox{\boldmath
$v_{i}$}$-$\mbox{\boldmath $v_{j}$}, \mbox{\boldmath
$r_{ij}$}=\mbox{\boldmath $r_{i}$}$-$\mbox{\boldmath $r_{j}$},
$\alpha$ and $\beta$ are constants, $c_{i}$ is the speed of sound at
the position of particle {\em i}, and
$\overline{c}_{ij}$=$(c_{i}+c_{j})/2$ and
$\overline{\rho}_{ij}$=$(\rho_{i}+\rho_{j})/2$. In the simulations
presented in this paper we have used $\alpha$=1, $\beta$=2 and
$\eta^{2}$=$10^{-2}$.

The form of \mbox {\boldmath $F_{iH}$} , and in particular the
symmetrization of the pressure gradient term, ensures that forces
between pairs of particles are symmetric, and so linear and angular
momentum are automatically conserved. The viscosity is introduced with
a term linear in the velocity differences, which produces shear and
bulk viscosity, and a quadratic term to handle high Mach numbers in
shocks, analogous to the Von Neumann--Richtmyer artificial viscosity
used in finite--difference methods~(\cite{richt}).

The value $u_{i}$ of the thermal energy per unit mass is evolved in
time according to the the first law of thermodynamics, taking into
account the contribution from the viscous terms. The symmetrized form
is:
\begin{eqnarray}
\frac{du_{i}}{dt} = \frac{1}{2} \sum_{j} m_{j} \left(
2\frac{\sqrt{P_{i}P_{j}}}{\rho_{i}\rho_{j}} +\Pi_{ij} \right)
\mbox{\boldmath $v_{ij}$} \cdot \mbox{\boldmath
$\nabla$}_{i}W_{ij}. \label{eq:dudt}
\end{eqnarray}

The thermal energy equation is integrated by essentially using a
two--step procedure~(\cite{HK}) in order to preserve accuracy. This is
necessary because the heating rate depends on the pressure, and via
the equation of state, on the thermal energy itself.

Individual smoothing lengths $h_{i}$ are adjusted so that every
particle has an approximately constant number $\nu$ of hydrodynamical
neighbors, i.e., those for which $r_{ij}/h_{ij} \leq 2$. This ensures
that adequate spatial resolution is maintained and that the level of
accuracy of the interpolations remains uniform throughout the
fluid. If $h_{i,n}$ is the smoothing length for particle {\em i} at
step {\em n}, the value at step $n+1$ is found according
to~(\cite{HK}):
\begin{eqnarray}
h_{i}^{n+1}=\frac{h_{i}^{n}}{2}
\left[1+\left(\frac{\nu}{\nu^{n}}\right)^{\frac{1}{3}}\right]
,\label{eq:dhdt}
\end{eqnarray}
where $\nu^{n}$ is the number of hydrodynamical neighbors at step {\em
n}. We have found that a value of $\nu$=64 yields adequate sampling
of the fluid without requiring excessive computation time.

Gravitational force calculations constitute the most time--consuming
part of an $N$--body code. A direct calculation requires $O(N^{2})$
operations, and this becomes prohibitive as the number of particles is
increased above a few hundred. In order to make this process more
efficient we have incorporated a binary tree structure~(\cite{HK})
into our code to obtain gravitational forces. This makes the number of
required operations of $O(N\log N$), thus allowing for simulations
with several thousand particles. The tree structure is built by
pairing two particles into a new pseudo--particle, called a node, if
they are mutual nearest neighbors, and proceeding with this pairing
over all particles and newly created nodes until there is only one
node, at the center of mass of the particles being considered. This
hierarchical association makes up the binary tree which we use to
compute the forces (for a detailed description see~\cite{Benz}). The
force on any particle can then be obtained by adding up contributions
from particles close enough to be considered individually and by more
distant aggregations, which can be described in terms of their
multipolar expansion up to the quadrupole term. To decide if an
aggregation of particles must be resolved to obtain the force on
particle {\em i}, we consider the ratio $s/d$, where {\em d} is the
distance from particle {\em i} to the aggregate and {\em s} is its
size. If $s/d \geq \theta$, where $\theta$ is a dimensionless
tolerance parameter, the aggregate is resolved further. If the
converse is true, the contribution to the force is obtained from the
quadrupole expansion. We adopt $\theta$=0.8, which provides an
accuracy for the force of better than one part in $10^{2}$. However,
the aggregate is resolved further if it contains any individual
particles which may be hydrodynamical neighbors for particle {\em
i}. The subroutine corresponding to this approach simultaneously
provides \mbox{\boldmath $F_{iG}$} and the hydrodynamical neighbors
for every particle that are used to obtain {\boldmath $F_{iH}$} and
$du_{i}/dt$ in equations~(\ref{eq:defhydrof}) and (\ref{eq:dudt}). Our
calculations of the gravitational forces are completely Newtonian.

We use a leapfrog algorithm accurate to second order to integrate the
equations of motion. The time step is adaptive, and is taken to
satisfy a combination of the Courant criterion for stability and a
further restriction on the maximum change allowed in velocity for any
particle during one time step to conserve accuracy. We
follow~\cite{Monaghan89} and set the time step to be $\Delta
t$=$\min{(\Delta t_{1}, \Delta t_{2})}$ with
\begin{eqnarray*} 
\Delta t_{1}= \min_{i}(h_{i}/\dot{v}_{i})^{1/2} \label{eq:step1}
\end{eqnarray*}
\begin{eqnarray*}
\Delta t_{2}=0.15 \min_{i} \left( \frac{h_{i}}{c_{i}+1.2 \alpha
c_{i}+1.2 \beta \max_{j}(\mu_{ij}) } \right) \label{eq:step2}
\end{eqnarray*}

One of the main objectives of the present work is to investigate the
gravitational waves emitted by coalescing binaries. In our Newtonian
simulations we calculate the waveforms of the emitted waves and the
total luminosity in the quadrupole approximation. The amplitude of the
retarded wave at time {\em t} and a distance $r_{0}$ away from the
source is given by ~(\cite{MTW})
\begin{eqnarray*}
h^{TT}_{jk}=\frac{2}{r_{0}} \frac{G}{c^{4}}
\Ibardd_{jk}\,^{TT}\left(t-\frac{r_{0}}{c}\right) ,\label{eq:htt}
\end{eqnarray*}
\begin{eqnarray*}
\mbox{where}\
\Ibar_{jk}\,^{TT}=P_{jn}\Ibar_{nm}P_{mk}-P_{jk}P_{lm}\Ibar_{lm}/2
\label{eq:Itt}
\end{eqnarray*}
is the transverse traceless part of the reduced quadrupole moment,
\begin{eqnarray*}
\Ibar_{jk}=I_{jk}-\delta_{jk}I ,\label{Itt2}
\end{eqnarray*}
\begin{eqnarray*}
\mbox{with}\ I_{jk}= \int \rho x_{j} x_{k} d\mbox{\boldmath$r$}
\label{eq:defI}
\end{eqnarray*}
\begin{eqnarray*}
\mbox{and}\ I=\sum_{j} I_{jj}. \label{eq:TraceI}
\end{eqnarray*}

The $P_{ij}$ are the components of the projection operators onto the
plane transverse to the radial direction: \( P_{ij}= \delta_{ij} -
n_{i}n_{j} \) with \(n_{i}=x_{i}/r_{0} \).

The luminosity of the gravitational waves at a distance {\em r} from
the source is
\begin{eqnarray}
L_{GW}=\frac{dE}{dt}=\frac{1}{5} \frac{G}{c^{5}} \left \langle
\frac{d\Ibardd_{jk}}{dt} \frac{d\Ibardd_{jk}}{dt} \right
\rangle .\label{eq:gravlum}
\end{eqnarray}

We use the method of~\cite{Finn} and obtain the formula applicable to
SPH for the second derivative of the quadrupole moment~(\cite{RS92}):

\begin{eqnarray*}
\ddot{I}_{fluid}^{jk}=\sum_{i} m_{i} \left( 2 v_{i}^{j} v_{i}^{k} +
\frac{2P_{i}}{\rho_{i}} \delta^{jk} + x_{i}^{k} g_{i}^{j} + x_{i}^{j}
g_{i}^{k}\right). \label{eq:nsddotI}
\end{eqnarray*}
The summation is over the SPH particles and the superscripts indicate
the cartesian components. We identify the gravitational acceleration
on particle {\em i} as \mbox{\boldmath $g_{i}$}. The term
$\ddot{I}_{fluid}^{jk}$ only includes the contribution from the SPH
particles contained in the neutron star. We add the terms arising from
the presence of the black hole as a point mass contribution:
\begin{eqnarray*}
\ddot{I}_{BH}^{jk}= M_{BH} \left( 2 v_{BH}^{j} v_{BH}^{k} + x_{BH}^{k}
g_{BH}^{j} + x_{BH}^{j} g_{BH}^{k}\right), \label{eq:bhddotI}
\end{eqnarray*}
to obtain the total value of $\ddot{I}$ as
\begin{eqnarray*}
\ddot{I}^{jk}=\ddot{I}_{BH}^{jk}+\ddot{I}_{fluid}^{jk}
.\label{eq:totalddotI}
\end{eqnarray*}

The gravitational waveforms are thus obtained directly from dynamical
and hydrodynamical variables of the system, and one numerical
differentiation of $\ddot{I}^{jk}$ is required to obtain the
luminosity.

The amplitudes of the waveforms for an observer a distance $r_{0}$
away along the axis of rotation of the binary are given by
\begin{eqnarray}
r_{0}h_{+}=\frac{G}{c^{4}}\left(\Ibardd_{xx}-\Ibardd_{yy}\right),
\label{eq:h+}
\end{eqnarray}
\begin{eqnarray}
r_{0}h_{\times}=2 \frac{G}{c^{4}}\Ibardd_{xy} .\label{eq:hx}
\end{eqnarray}

\newpage

\begin{deluxetable}{lccccc}
\tablewidth{0pt} \tablecaption{Comparison of results for the
coalescence of two identical polytropes with a stiff equation of state
($\Gamma=3$).
\label{RSLK}} \tablehead{ & \colhead{$(r_{0}R/M^{2})h_{max}$\tablenotemark{a}} &
\colhead{$(r_{0}R/M^{2})h_{final}$\tablenotemark{a}} &
\colhead{$(R/M)^{5}(L_{max}/L_{0})$\tablenotemark{b}} & \colhead{$\rho_{c}$} &
\colhead{$M_{halo}/M_{tot}$} } 
\startdata 
This work & 2.2 & 0.2 & 0.39 & 0.44 & 0.16 \nl
RS        & 2.2 & 0.2 & 0.37 & 0.40 & 0.18 \nl 
\enddata

\tablenotetext{a}{The peak and final gravitational radiation
amplitudes at a distance $r_{0}$ from the source (see
equations~[\ref{eq:h+}] and~[\ref{eq:hx}]).}

\tablenotetext{b}{See
equation~(\ref{eq:gravlum}). $L_{0}=c^{5}/G=3.59\times
10^{59}$~erg~s$^{-1}$}

\end{deluxetable}

\newpage

\begin{deluxetable}{ccc}

\tablewidth{0pt} \tablecaption{Critical Separations for black
hole--neutron star binaries. \label{critical}} 
\tablehead{ \colhead{$q$}
& \colhead{$r_{min}$} & \colhead{$r_{RL}$}} 
\startdata 
1.00 & 2.82 & 2.78 \nl 
0.80 & 2.97 & 2.94 \nl 
0.31 & \nodata & 3.76 \nl 
\enddata
\end{deluxetable}

\newpage

\begin{deluxetable}{cccccccc}

\tablewidth{0pt}
\tablecaption{Summary of results for the coalescence of two neutron stars and a black hole with a neutron star\tablenotemark{a}. \label{summary}}
\tablehead{ \colhead{} & \colhead{$q_{i}$} & \colhead{$q_{f}$} & \colhead{Torus} & \colhead{$(r_{0}R/M^{2})h_{max}$}  & \colhead{$(r_{0}R/M^{2})h_{final}$} & \colhead{$(R/M)^{5}(L_{max}/L_{0})$} & \colhead{$M_{core}$}}
\startdata
NS--NS & 1.000 & 1.000 & no & 2.20 & 0.20 & 0.390 & 0.82 \nl
BH--NS & 1.000 & 0.190 & yes & 1.75 & 0.55 & 0.145 & 0.3070 \nl
BH--NS & 0.800 & 0.360 & no &  1.93 & 1.25 & 0.160 & 0.5950 \nl
BH--NS & 0.310 & 0.306 & no & 3.59  & 3.48 & 0.425 & 0.9896 \nl
\enddata
\tablenotetext{a}{Notation is the same as in Table~\ref{RSLK}}
\end{deluxetable}

\end{document}